# Interplay of topology and antiferromagnetic order in two-dimensional van der Waals crystals of $(Ni_xFe_{1-x})_2P_2S_6$

N. Khan[1,*], D. Kumar[1], V. Kumar[1], Y. Shemerliuk[2], S. Selter[2], B. Büchner[2,4], K. Pal[3], S. Aswartham[2] and Pradeep Kumar[1,†]

[1] *School of Physical Sciences, Indian Institute of Technology Mandi, Mandi-175005, India*
[2] *Leibniz-Institute for Solid-state and Materials Research, IFW-Dresden, 01069 Dresden, Germany*
[3] *Department of Physics, Indian Institute of Technology Kanpur, Kanpur- 208016, India*
[4] *Institute of Solid State and Materials Physics and Würzburg-Dresden Cluster of Excellence ct.qmat, Technische Universität Dresden, 01062 Dresden, Germany*

## Abstract

Mermin-Wagner theorem forbid spontaneous symmetry breaking of spins in one/two-dimensional systems at finite temperature and rules out the stabilization of this ordered state. However, it does not apply to all types of phase transitions in low dimensions such as topologically ordered phase rigorously shown by Berezinskii-Kosterlitz-Thouless (BKT) and experimentally realized in very limited systems such as superfluids, superconducting thin films. Quasi two-dimensional van der Waals magnets provide an ideal platform to investigate the fundamentals of low-dimensional magnetism. We explored the quasi two-dimensional (2D) honeycomb antiferromagnetic single crystals of $(Ni_xFe_{1-x})_2P_2S_6$ (x = 1, 0.7, 0.5, 0.3 & 0) with varying spins ($1 \leq s \leq 2$) using in depth temperature dependent Raman measurements supported by first-principles calculations of the phonon frequencies. As a function of doping, a tunable transition from paramagnetic to antiferromagnetic ordering is shown via phonons reflected in the strong renormalization of the self-energy parameters of the Raman active phonon modes. An anomalously broad magnetic continuum attributed to two-magnon excitations is observed and its coupling with the phonons is revealed in the observation a Fano line asymmetry. Interestingly, the two-magnon continuum is observed only for the finite doping (x ≠ 0) understood invoking underlying nature of insulator these materials belongs to, i.e. exchange interaction between transition metals via surrounding ligands (sulphur) and the resonance involving phonon modes associated with the ($P_2S_6$) cage. The extracted exchange parameter ($J$) is found to vary by ~ 100 % with increasing the value of doping, ranging from ~ 6 meV (for x = 0.3) to 13 meV (for x = 1). Quite surprisingly, we also observed renormalization of the phonon modes much below the long-range magnetic ordered temperature attributed to the topological ordered state, namely the BKT phase, which is also found to change as a function of doping. The extracted critical exponent of the order-parameter [spin-spin correlation length, $\xi(T)$] evince the signature of topologically active state driven by vortex-antivortex excitations.

*nasarukhan736@gmail.com
†pkumar@iitmandi.ac.in

# 1. Introduction

Transition metal phosphorus trisulfide ($TM_2P_2S_6$; TM = Mn, Fe, Co, Ni) compounds are layered quasi 2D materials, which are van der Waals stacked and have emerged as a promising class of materials for potential applications as well as exploring rich fundamental physics due to their complex magnetic properties [1-5]. The family of $TM_2P_2S_6$ offers a series of isostructural van der Waals layered compounds, all exhibiting antiferromagnetic order in bulk material but with different magnetic anisotropies. For example, $Mn_2P_2S_6$ is a Heisenberg antiferromagnet below Néel temperature of $T_N$ ~ 78 K, $Fe_2P_2S_6$ is an Ising antiferromagnet below $T_N$ ~123 K and $Ni_2P_2S_6$ exhibits an antiferromagnetic state below $T_N$ ~ 155 K, which can be described by an anisotropic Heisenberg model or the XXZ model [3,5-6]. In these systems magnetic properties can be tuned by chemical substitution, i.e., by gradually substituting Ni with Fe in $Ni_2P_2S_6$, which makes it to tune the magnetic anisotropy in the system. This will make it feasible to investigate the interplay between magnetic anisotropy, magnetic interactions and spin fluctuations in these magnetic systems. The work by Masubuchi et al. [7] on $(Mn_{1-x}Fe_x)_2P_2S_6$ shows the feasibility of substitution between two magnetic $TM_2P_2S_6$ compounds while preserving their intrinsic nature, which is crucial for a gradual evolution of the physical properties of material as a function of substitution. However, a breakdown in the long-range magnetic order and a spin-glass state around $(Mn_{0.5}Fe_{0.5})_2P_2S_6$ occurs instead of a gradual evolution between the magnetic characteristics of both parent compounds. The work by Shemerliuk et al. [8] on the substitution series $(Mn_{1-x}Ni_x)_2P_2S_6$ shows long-range antiferromagnetic order but with non-monotonous changing the Néel temperature and the switches of the magnetic anisotropy from out-of-plane to in-plane as a function of composition. On the other hand, a work by Rao et al., on the polycrystalline samples of $(Fe_{1-x}Ni_x)_2P_2S_6$ [9], demonstrated that the long range magnetic order exists throughout the series, hence this series proved to be a promising candidate for tuning the magnetic anisotropy. Exploitation of spin dependence of Raman process makes it possible to study the magnetic orders via Raman scattering. This is a powerful technique to study the various properties of materials such as strain [10], defect and doping [11], spin-phonon coupling [12-15], electron-phonon coupling [16-19] and effect of layer numbers [20,21] and magnetic excitations [22-25].

A unique property of these systems is that both their magnetic and crystal structures is quasi 2D, which set them apart from other 2D magnetic systems with three dimensional crystal structure. Mermin-Wagner in their seminal work showed that isotropic spins can't have long-range order in one/two-dimensions at finite temperature [26] by breaking the continuous global symmetry, clearly in contrast with the spin systems of these layered van der Waals compounds which orders at finite temperature, e.g. $Ni_2P_2S_6$ has $T_N \sim 155K$. With further lowering the temperature these systems are expected to go into the XY state, suggesting without any long-range ordering within the Mermin-Wagner theorem though this may be captured by the binding of magnetic vortex-antivortex pairs. Specifically, a topologically active phase transition may take place at non-zero temperature due to binding of vortex-antivortex pairs, namely the BKT transition [27]. Below the BKT transition temperature ($T_{BKT}$) spins can order with in the 2D plane at finite temperature even without the presence of 3D interaction. This distinct ordering of spins has the potential to affect the lattice degrees of freedom, as spins reside on the lattice and interfere in the atomic movement, and may reflect their ordering via renormalization of the self-energy parameters of the phonon modes. This renormalization may be captured by the Raman measurements in addition to the renormalization due to the antiferromagnetic ordering at higher temperature as Raman scattering is a powerful tool to understand the underlying magnetic texture in 2D magnetic materials. We note that evidence of the possible BKT transition in some members of the $TM_2P_2S_6$ family have been reported in the literature using different studies [28-30].

In this work, we report an in-depth temperature dependent Raman scattering measurements on quasi two-dimensional (2D) honeycomb antiferromagnetic single crystals of $(Ni_xFe_{1-x})_2P_2S_6$ ($x$ = 1, 0.7, 0.5, 0.3 and 0). We observed clear signature of antiferromagnetic ordering reflected in the renormalization of the self-energy parameters of the phonons as well as the integrated intensity, magnetic excitations (magnons) and the dynamic fluctuations extracted by the dynamic Raman susceptibility. This is followed by low temperature topological phase transition marked by the distinct changes in the phonon modes and the extracted critical exponent of the order-parameter using spin correlation length from our Raman measurements. This low temperature transition is attributed to the BKT phase.

## 2. Result and Discussion

### 2.1 Raman scattering for $(Ni_xFe_{1-x})_2P_2S_6$

At the room temperature and ambient conditions, bulk $Ni_2P_2S_6/Fe_2P_2S_6$ belongs to the point group $C_{2h}$ and space group $C2/m$ [#12]. Both crystalizes in the monoclinic structure, where all the Ni/Fe atoms are arranged in a hexagonal structure and each Ni/Fe atom is surrounded by six S atoms with trigonal symmetry. Two $P$ atoms are connected with six $S$ atoms, these $S$ and $P$ atoms are covalently bonded which makes $(P_2S_6)^{-4}$ anion complex (see supplementary Fig. S1). The unit cell of bulk $Ni_2P_2S_6$ and $Fe_2P_2S_6$ give rise to 30 phonon branches at $\Gamma$ point of the Brillouin zone with the following irreducible representations $8A_g + 6A_u + 7B_g + 9B_u$. Out of these 30 phonon branches 15 are Raman active ($\Gamma_{Raman} = 8A_g + 7B_g$), 12 are infrared active ($\Gamma_{IR} = 5A_u + 7B_u$), and 3 corresponds to the acoustic phonons $\Gamma_{Acoustic} = A_u + 2B_u$ [31]. Figure 1 shows the Raman spectrum for complete series of $(Ni_xFe_{1-x})_2P_2S_6$ at 4 K. The spectra are fitted with a sum of Lorentzian functions in order to extract the self-energy parameters i.e. mode frequency (ω) and full width at half maximum (FWHM)/linewidth of the individual mode. For convenience, we have labeled the observed modes as P1-P10, S1-S4 and J1-J2, see Fig. 1 (a-e). Frequency of the observed modes at 4 K and corresponding symmetry assignment are listed in Table-S1. Symmetry assignment of the modes is done in accordance to the previous reports and our DFT based calculations [31,32]. It should be noted that we observed the Fano asymmetry for the case of mode P8, see inset in Fig. 1(a), and interestingly this asymmetric nature of P8 mode is observed only for the case of $Ni_2P_2S_6$ which will be discussed later in details. We also observed a very broad peak attributed to the two-magnon excitation in the frequency range of 300 to 750 $cm^{-1}$, and quite surprisingly its intensity is observed to decrease with increasing Fe doping, see Fig. 1, and it completely vanishes for the case of $Fe_2P_2S_6$.

In $TM_2P_2S_6$, the vibrational phonon modes may be divided into external modes (low frequency vibrational modes below ~ 250 $cm^{-1}$), which appears primarily due to the vibrations of heavy metal ions and phosphorus atoms, and internal modes (high frequency vibrational modes above ~ 250 $cm^{-1}$) which generally appears due to the vibrations of $(P_2S_6)^{-4}$ structure [31]. Since the high frequency vibrational modes appear due to the vibrations of $(P_2S_6)^{-4}$ structure, therefore these modes are expected to be identical in all $TM_2P_2S_6$ materials. We observed mode P1 for the case of $Ni_2P_2S_6$, $(Ni_{0.7}Fe_{0.3})_2P_2S_6$, $(Ni_{0.5}Fe_{0.5})_2P_2S_6$ and $(Ni_{0.3}Fe_{0.7})_2P_2S_6$ and this may be attributed to the vibrations of Ni atoms or combination of Ni/Fe atoms. Further, we also observed the reduction in intensity of P1 mode with increase in Fe doping, and for the case of $Fe_2P_2S_6$ we did not observe any signature of P1 mode. Mode P2 appears throughout the series and is due to vibrations of the transition metal atom [33]. Modes P3-P6 and P8 are also observed in all the compounds, see Fig. 1 (a-e), and these modes may be attributed to the vibrations of $(P_2S_6)^{-4}$ structure. A relatively broad and weak mode (P7) centered near ~ 445 cm$^{-1}$ is assigned as infrared active mode with $A_u$ and/or $B_u$ symmetry [31]. We also observed mode P9 and P10 centered near ~ 590 cm$^{-1}$ for all the compounds of series except $Fe_2P_2S_6$. We also observed few additional modes labelled as S1-S4 for the case of doped samples only i.e. $(Ni_{0.7}Fe_{0.3})_2P_2S_6$, $(Ni_{0.5}Fe_{0.5})_2P_2S_6$ and $(Ni_{0.3}Fe_{0.7})_2P_2S_6$, see Fig. 1 (b-d). These S1-S4 modes may have their origin due to the combined effect of Ni and Fe atoms, so these modes might have appeared nearby P1 and P2 modes which are solely attributed to the Ni or Fe atom. These may also have their origin in the modes away from the $\Gamma$ points, which may become active in the doped samples owing to broken symmetry as it relaxes the zero-momentum constraint for modes to be Raman active. Also, these may be due to possible change in crystal symmetry from *C2/m* to *P2/m*, *P2/c* or *C2* due to doping which may leads to more Raman active phonon modes [34]. All these new modes (S1-S4) in the doped samples appears below ~ 250 cm$^{-1}$ suggesting their origin involving vibrations of the TM atoms. In addition to the observed mode P1-P10 and S1-S4, we also observed a very interesting low frequency feature only for the case of $Fe_2P_2S_6$, labelled as *J1*. In addition, we note that temperature dependent Raman measurements on the members of $TM_2P_2S_6$ family have uncovered rich optically active features especially below the long-range magnetic ordering temperature. The origin of these features was attributed to multiple factors such as spin-spin and spin-phonon coupling, Brillouin zone folding, and quantum interference of the discrete phononic state with the electronic excitations owing to the spin splitting of the band structure [31,35-38]. We note that for the case of $Ni_2P_2S_6$, similar to Mn$_2$P$_2$S$_6$, the underlying chemical and magnetic unit cell are same, as a result no new zone folded phonon modes are expected in line with our observations. On the other hand, for the case of Fe$_2$P$_2$S$_6$, magnetic ordering result into doubling of magnetic unit cell for both in-plane and out-of-plane direction and it may result into new zone folded modes below $T_N$. In particular $M$ point in the Brillouin zone folds into $\Gamma$ point below $T_N$. We note that for the case of Fe$_2$P$_2$S$_6$, a broad mode near ~ 100 cm$^{-1}$ is reported in the literature similar to $J1$ mode in our results, which splits into 3 modes below $T_N$ along with the emergence of a new mode near ~ 120 cm$^{-1}$ attributed to one-magnon excitations [38-40].

## 2.2 Temperature dependence of the Phonon modes

Figure 2 (a) and 2 (b) illustrate the self-energy parameters (i.e. mode frequency and FWHM) of the prominent phonon modes as a function of temperature for the end members of the series i.e. Ni$_2$P$_2$S$_6$ and Fe$_2$P$_2$S$_6$ respectively. Temperature dependence of the mode frequency ($\omega$) and FWHM for the doped cases (x = 0.7, 0.5 & 0.3) are shown in the supplementary Fig. S3 and S4. The following observations can be made: (i) The frequencies of all the observed prominent phonon modes (some modes are not shown here) P1, P3, P4, P6, P7, P9 for Ni$_2$P$_2$S$_6$ and P2, P3, P4, P5, P6 for Fe$_2$P$_2$S$_6$ show normal behaviour i.e., mode softening with increasing temperature from ~$T_N$ (Néel Temperature) to 330 K; may be understood within the anharmonic phonon-phonon interaction model (discussed in detailed in supplementary section-I). (ii) Linewidth of the modes P1, P4, P6, P9 (for Ni$_2$P$_2$S$_6$) and P2, P4, P5, P6 (for Fe$_2$P$_2$S$_6$) also show normal behaviour i.e., increasing linewidth with increasing temperature above $T_N$. (iii) Interestingly mode frequency (except in the temperature range of $T_{BKT}$ to $T_N$) and FWHM of the J1 mode (Fe$_2$P$_2$S$_6$) show the anomalous behaviour i.e., frequency increases and FWHM decreases with increasing temperature. Similar anomalous behaviour is also shown by the S1 mode (see supplementary Fig. S4) for three members of the series (Ni$_x$Fe$_{1-x}$)$_2$P$_2$S$_6$ i.e., x=0.7, 0.5 & 0.3.

Near the vicinity of the magnetic ordering temperature correlation between spins starts building up and these correlation starts interfering with the dynamics of underlying lattice vibrations, generally reflected in renormalization of the phonon's self-energy parameters in the ordered phase. We observed anomalous behaviour of the frequency, FWHM as well as intensity (shown in the supplementary Fig. S5) in the vicinity of paramagnetic to antiferromagnetic phase transition for $Ni_2P_2S_6$ ($T_N \sim 155$ K) and $Fe_2P_2S_6$ ($T_N \sim 123$ K). We note that in the vicinity of these temperatures, there is a clear change in the slope of frequencies and FWHM suggesting entering into the long-range magnetically ordered phase. Similar changes are also seen for the doped samples of the series (see supplementary Fig. S3). Frequency of P1 and P2 (not shown here) modes of $Ni_2P_2S_6$ remains nearly constant while for P3, P4, P6 and P9 mode; it increases slowly from ~155 K to ~ 55 K. Frequency of P7 mode increases rapidly as the temperature decrease from ~155 K to ~ 55 K in comparison to the temperature range of ~ 155 K to 330 K. For the case of $Fe_2P_2S_6$ frequency of P2, P3, P4, P5, P6 modes increases rapidly between ~123 K to ~30 K. For both these systems there is a clear change in the slope in FWHM in the vicinity of aforementioned temperatures (~155 K for $Ni_2P_2S_6$ and ~123 K for $Fe_2P_2S_6$). Linewidths of P1, P4, P7, P9 for the case of $Ni_2P_2S_6$ and P2 and P6 for the case of $Fe_2P_2S_6$ shows a cusp in the vicinity of these temperatures i.e. linewidth first increases and then decreases in a short temperature window; clearly reflecting the transition from paramagnetic to antiferromagnetic phase. As the system begins to enter into the long-range ordered phase, we observed that phonon modes frequencies show clear renormalization, where some modes showed small changes, while others exhibited prominent changes. This variation in the phonon modes renormalization may be captured via understanding symmetry and magnitude of the underlying lattice vibrations [25]. Also, with the building up of spin-spin correlation, more decay channel will be available for the phonons and as a result generally linewidth is expected to increase but the magnitude of renormalization is affected by the dynamics of the phonons involved. We clearly see cusp in mode linewidths in the vicinity of the temperature at the interface of para to antiferromagnetic transition. Once deep into the long range ordered phase fluctuation die down and linewidth also decreases with further decreasing the temperature. Using the anomalies in phonon self-energy parameters in the prominent phonon modes, we extracted the antiferromagnetic transition temperature ($T_N$). Figure 3 shows the evolution of $T_N$ as a function of increasing Ni percentage in $(Ni_xFe_{1-x})_2P_2S_6$ extracted from our Raman measurements. We found that $T_N$ extracted using Raman data increases gradually as one moves from $Fe_2P_2S_6$ to $Ni_2P_2S_6$ matching quite nicely with the data extracted from the magnetization measurements on these samples [41]. Table-S2 shows the extracted value of $T_N$ from our Raman measurement for all members of the series.

Quite surprisingly, we observed signature of another ordering much below the respective antiferromagnetic transition temperature reflected in the renormalization of the phonon modes self-energy parameters. In particular, we observed anomalous behaviour in the mode frequency and FWHM. In the temperature range of ~ 55 K to 4 K, frequency for all the shown modes of $Ni_2P_2S_6$ takes a sharp upturn; on the other hand, for the case of $Fe_2P_2S_6$, it shows a clear downturn in the temperature range of ~ 30 K to 4 K. There is a clear change in slope of the FWHM at these temperatures for both the samples, see Fig. 2. We note that a few systems with 2D magnetic nature have been shown to exist, which are mainly governed by the Ising or Heisenberg type interactions [40,42-43] including members of this $TM_2P_2S_6$ family. It has also been shown rigorously that in XY-like 2D systems a special topological transition may exist reflected by the transition from bound vortex-antivortex pairs into free vortices [27]. Above this topological transition temperature, these vortex-antivortex pairs deconfine into a plasma of mobile vortices, and the same is reflected via an exponential decay of the correlation length with temperature. Systems describable with in the XY model (having single ion anisotropy and with weak interlayer interaction) are conjectured to be suitable candidates for observing this topologically active transition named after the scientists as BKT transition. The XY type nature of the spin dimensionality, where spins are constrained to rotate in plane of the lattice, has been experimentally shown even for the case of bulk $(Ni/Mn)_2P_2S_6$ [28,30,43-44], and suggested the possible existence of vortices, a hallmark of topological transition, in these systems.

Therefore, the observed renormalization of the phonon modes much below the antiferromagnetic transition temperature suggest its origin in the topological ordering of the spins (BKT transition) with no conventional symmetry breaking which arises from the bonding of vortex-antivortex pairs at low temperature. It has been reported that these systems show the BKT transition at low temperature due to the binding of magnetic vortex-antivortex pairs. So, it may be considered that in the $Ni_2P_2S_6$ and $Fe_2P_2S_6$ spins may begin to constrained in a plane at the low temperature and resultantly these transforms into XY system and show the BKT type of topological ordering [27,29]. We also extracted $T_{BKT}$ from our data using the renormalization in the phonon parameters and is shown in Fig. 3 as a function of Ni doping for $(Ni_xFe_{1-x})_2P_2S_6$ and is found to increase as one moves from $Fe_2P_2S_6$ to $Ni_2P_2S_6$. We wish to note that ours is the first-report where possible signature of topological ordering on the phonon renormalization are identified. It has been suggested that this topological active BKT transition is destroyed by the external magnetic field perpendicular to the sample surface (along the *c*-axis) [29]. It will be interesting to probe the phonon dynamics in these systems under magnetic field and shed more light on the nature of this topological ordering.

## 2.3 Two-Magnon Scattering

Magnetic two-dimensional van der Waals systems have potential to tune with external parameters. The recent theoretical prediction of topological magnons in 2D systems has gained lot of attention [45-46]. Hence $TM_2P_2S_6$ is an ideal family to investigate low temperature magnetic excitations with various tuning parameters such as doping [47-48]. Dynamics of coupled spin and lattice degrees of freedom play an important role in varied phenomena such as controlling spin ordering via phonons, driving multiferroic polarization, spintronic, to name a few [49-50].

In order to understand the dependency of temperature and chemical tunning on the antiferromagnetic order, we now focus to understand the distinctive features of magnetic excitations observed at low temperatures. Figure 4(a) shows the Raman spectra of $(Ni_xFe_{1-x})_2P_2S_6$ series at low temperature in the spectral range between 50 - 750 cm$^{-1}$. At low temperature we observed a broad peak, centred at ~ 530 cm$^{-1}$ for the case of $Ni_2P_2S_6$ (see top panel of Fig. 4(a)). We attribute this broad continuum to the two-magnon (2M) scattering in line with the reports in the literature [32,51]. We note that at low temperature this continuum is prominent for the case of $Ni_2P_2S_6$, but as percentage of Ni starts decreasing the 2M peak also starts weakning, both its intensity and FWHM decreases. Interestingly for the case of $Fe_2P_2S_6$ we did not observe any 2M signal, attributed to the absence or minimal effect of superexchange process, Mott-Hubbard insulating nature and resonance involving phonons; discussed in detail in the supplementary section-III. We also observed that with decreasing percentage of Ni, the 2M peak energy ($\Omega_0$) also decreases i.e. its peak start shifting towards the lower frequency; this also hints the possible decreasing trend in the spin-spin exchange interaction parameter ($J$) with increasing Fe content.

It is observed that for all the samples 2M intensity is growing continuously with decreasing temperature below $T_N$ (see supplementary Fig. S6). These results are generally expected for a typical antiferromagnetic material, where two-magnon signal gets weaker, redshifted, and broadened as the corresponding spectral weight is shifted towardes quasi elastic scattering part with increasing temperature [52-54]. Figure 4(b) and (c) show temperature evolution of the peak frequency and FWHM of this 2M mode for $(Ni_xFe_{1-x})_2P_2S_6$, respectively. It can be observed that its frequency is decreasing continuously with increasing temperature for all the compounds. For the case of $Ni_2P_2S_6$, FWHM increases with increasing temperature but for other doped systems the trend is different. FWHM decreases with increasing temperature for x = 0.7 , 0.3 and for x = 0.5 it first increases with increasing temperature and then shows a broad maxima and after that it again starts to increase with increasing temperature. As temperature increases the 2M signal becomes weaker but show weak signal even well above $T_N$ till ~ 200 K, (see the supplementary Fig. S6). Figure 4(d) shows the intensity plot, it is clear that intensity of 2M start building up from around 200K and it rises sharply below respective $T_N$.

From the energy of the two-magnon peak the nearest neighbour exchange parameter ( $J$ ) may be estimated. Within the simplestic approximation for a typical AFM system, if spin deviations are created on the adjacent sites, the two-magnon energy may be expressed $E_0 = J(2Sz - 1)$ [55], where $S$ is spin of the magnetic site $Ni_x/Fe_{1-x}$, which is evaluated by taking average as $S = xS_{Ni} + (1 - x)S_{Fe}$, $S_{Ni}/S_{Fe}$ is spin on $Ni^{2+}/Fe^{2+}$ site and z is the number of nearest neighbours to that magnetic site (here z = 3). Though the actual magnetic structure is quite complex being three different interaction parameters namely $J_1$, $J_2$ and $J_3$; but the two of these are an order of magnitude smaller than the third one. In the paramagnetic phase above $T_N$ there is no preference between AFM and ferromagnetic (FM) bonds between Ni atoms. However below $T_N$, this equivalency will break. For the 1st/2nd/3rd nearest neighbour Ni atoms having $J_1/J_2/J_3$ interaction parameter there are 2FM/1AFM, 2FM/4AFM and 3AFM bonds, respectively (see supple. Fig. S1 (d)). It has been advocated that $J_3$ is an order of magntidue larger than $J_1$ , and $J_2$ is negligible as compared to $J_3$ as well as $J_1$ [51] in line with the increased number of AFM bonds. Keeping the above reported facts about the strength of the intercation parameters, therefore, for simplicity one may considered effectively only one interaction parameter below $T_N$. Using $\omega$ = 532.3 cm⁻¹, 434.7 cm⁻¹, 425.2 cm⁻¹, 428.1 cm⁻¹ for x = 1, 0.7, 0.5 & 0.3, respectively. The estimated values of nearest neighbour exchange parameters are found to be 13.2 meV, 7.9 meV, 6.6 meV, 5.8 meV for x = 1 [ $Ni_2P_2S_6$ ], x = 0.7 [ $(Ni_{0.7}Fe_{0.3})_2P_2S_6$ ], x = 0.5 [ $(Ni_{0.5}Fe_{0.5})_2P_2S_6$ ], x = 0.3 [ $(Ni_{0.3}Fe_{0.7})_2P_2S_6$ ], respectively. It could be seen that $J$ is decreasing with decreasing

Ni content (see supplementray Fig. S7 for 2M energy and $J$ as a function of Ni doping). We note that the estimated value of $J$ for $Ni_2P_2S_6$, is in quite good agreement with what has been reported in earlier studies [51].

We note that linewidth of the 2M peak is unusually large similar to those reported for the cuprates [56], incompatible within the conventional models [57]. The observed linewidth is 4-5 times higher than expected from the purely spin-wave predictions. Specifically, the ratio $(\gamma_0 / \Omega_0)$ of FWHM ($\gamma_0$) to the peak frequency ($\Omega_0$) at 4K for $Ni_2P_2S_6$ and other doped systems varies from 0.48 to 0.2 against the predicted 0.1 using the established theory for contribution of a single pair of AFM coupled spins. Our observation of unusual large linewidth of 2M hints the strong magnetostrictive interaction ($U = \sum_{i,j} \left| \frac{\partial J_{ij}}{\partial Q} \right|_{Q=0} (\vec{S}_i . \vec{S}_j) Q$; where $J$ is the exchange parameter and $Q$ is the generalized phonon coordinate) relying on the modulation of $J$ with the lattice vibrations and its affect also be visible in the phonon modes evolution. This interaction (U) leads to 2M-phonon coupling giving ways for magnon and phonon decay. We note that this magnetic continuum is coupled strongly to the phonon modes, whose scattering intensity interferes with it as reflected in the asymmetric Fano line shapes of the observed phonons, discussed in the next section.

## 2.4 Fano Resonance

Several features of the Raman spectra may yield crucial information about the underlying physics beyond the spectral peak position, FWHM and the overall line shape. Various interactions may affect the Raman line shape, which may shift it away from its conventional symmetrical Lorentzian shape. The asymmetric line shape reflects the interaction of the phonons with the underlying magnetic continuum, and it may be used to extract the important information to shed light on the nature of the coupling. Fano resonance is a result of purely quantum process and has its origin in the spin-dependent electron polarizability involving both spin-photon and spin-phonon coupling [58-61], with the coupling between discrete state (a phonon mode here) and continuum state (electronic or magnetic in nature) determining the degree of asymmetry.

We observed an asymmetric phonon mode P8 in the Raman spectrum of $Ni_2P_2S_6$. This asymmetry is attributed to the Fano resonance, which appears due to the coupling between magnetic continuum and a phonon mode. The P8 peak shows Fano nature only in case of $Ni_2P_2S_6$, this is probably due to the two-magnon peak shifting towards lower energy with increasing Fe percentage hence there is no coupling or negligible coupling between the magnetic continuum (two-magnon peak) and discreate phonon peak (P8 phonon mode) for x = 0.7, 0.5, and 0.3. We fitted this asymmetric mode P8 using Bright-Wigner Fano (BWF) function expressed as [32,58,61]:

$$I = I_0 \frac{[1 + 2(\omega - \omega_0)/(q\Gamma)]^2}{[1 + 4(\omega - \omega_0)^2/\Gamma^2]} \qquad (1)$$

where $\omega_0$ is the uncoupled phonon frequency, $\Gamma$ is the uncoupled linewidth and $1/q$ is the Fano asymmetry parameter. Figure 5(a) shows temperature evolution of the mode (P8) for $Ni_2P_2S_6$. It can be observed that below $T_N$ (~155 K) it shows significant asymmetry. Figure 5(d) shows the temperature evolution of Fano asymmetry parameter $(1/|q|)$. The Fano asymmetry parameter characterizes the coupling strength to the underlying continuum, as $1/q \rightarrow \infty$ represents a stronger coupling which causes the peak to be more asymmetric and in the case of $1/q \rightarrow 0$ it represents weak coupling which causes the peak to be symmetric and reduces to a Lorentzian line shape. We note that, with decreasing the temperature asymmetry increases continuously especially below $T_N$. It is also observed that as temperature increases beyond $T_N$ coupling strength parameter decreases and near room temperature it shows nearly symmetric nature. Figure 5(b) and 5(c) illustrate temperature evolution of the frequency and FWHM respectively. We note that in addition to the renormalization near $T_N$, we do see a sharp upturn in the frequency and FWHM below ~ 55 K, suggesting topologically ordered phase below this temperature.

## 2.5 Temperature Dependence of the Quasi-Elastic Scattering

We also analysed the magnetic fluctuations induced low frequency quasi elastic scattering (QES). Its temperature evolution and quantitative analysis may be used as good gauzing parameter for magnetic ordering. These QES signals are suppressed at low temperature due to the presence of the spin ordering in antiferromagnetic or ferromagnetic materials. For a quantitative analysis we follow the theory given by Reiter and Halley [62,63]. Intensity expression of the stokes Raman shifted QES may be given as [62-67]:

$$I(\omega) \propto \frac{\omega}{1 - e^{-\hbar\omega/k_B T}} \frac{C_m T D k^2}{\omega^2 + (Dk^2)^2} \qquad (2)$$

where $C_m$ is the magnetic specific heat at zero magnetic field, D is thermal diffusion constant ($D = K / C_m$) with magnetic contribution to thermal conductivity $K$. Raman response $\chi^{''}(\omega)$ for stokes scattering is obtained by dividing intensity by $[n(\omega)+1]$, where $n(\omega)$ is the Bose-Einstein factor. It reflects the dynamic properties of collective underlying excitations. Further to obtain Raman conductivity we divided Raman response by frequency ($\chi^{''}/\omega$). From equation (4) Raman conductivity may be expressed as:

$$\frac{\chi''(\omega)}{\omega} \propto C_m T \frac{Dk^2}{\omega^2 + (Dk^2)^2} \qquad (3)$$

Supplementary Fig. S8 (a-e) and Fig. S8 (f-j) show the QES and Raman response, respectively, for the $(Ni_xFe_{1-x})_2P_2S_6$ series. Raman conductivity is shown in the supplementary Fig. S9. To quantitatively understand the quasi elastic scattering, we evaluated the dynamic Raman succeptibility which is obtained from Raman conductivity using Kramers-Kronig relation [68], and is given as:

$$\chi^{dyn}(q=0,T) = \frac{2}{\pi} \int_0^\Omega \frac{\chi''(\omega,T)}{\omega} d\omega \qquad (4)$$

Dynamic Raman susceptibility is obtained by integrating the Raman conductivity to a finite frequency limit, the upper cut-off value is taken as 55 cm$^{-1}$, after which there is no significant change in Raman conductivity with increasing the Raman shift. Figure 6(a-e) shows the temperature evolution of the dynamic Raman susceptibility for all compounds of the series. It can be observed that for all the compounds dynamic Raman susceptibility remains independent or decreases slightly from the 330 K to ~ $T_N$. Below $T_N$ it decreases sharply because as the temperature decreases spin fluctuations are expected to suppress due to setting up of the long-range magnetic order. Above $T_N$, for Fe$_2$P$_2$S$_6$ $\chi^{dyn.}$ remains nearly constant and for other constituents of the series it increases slightly as it is expected in the paramagnetic phase that spins do not exhibit correlation so dynamic Raman susceptibility should remain nearly temperature independent. In the antiferromagnetic ordered phase at low temperature (~ 55 K) a slope change can be observed in the dynamic Raman susceptibility for the case of Ni$_2$P$_2$S$_6$. It may be due to BKT transition, as spin fluctuations are probably related to the bound vortex-antivortex pairs [27,69], which increases as temperature approaches the boundary of this transition at ~ 50 K.

Another quantifiable parameter is the spin correlation length, $\xi(T)$, which may be extracted from the Raman conductivity. $\xi(T)$ is obtained as inverse of the FWHM of the Lorentzian profile. We noted that Raman conductivity ($\chi''/\omega$) follow the Lorentzian shape spectral function. To fit it using a Lorentzian profile, we have taken the spectral range till 55 cm$^{-1}$ and we extrapolated this to 0 cm$^{-1}$ and taken the image of spectral range till -55cm$^{-1}$. Figure 6 (f, g) shows the temperature dependence of the extracted spin correlation length. We analysed the correlation length in terms of the BKT exponential scaling law for the 2D XY model, given as $\xi(T) \propto \exp(b[\sqrt{T_{BKT}/(T - T_{BKT})}])$ [27]. Here, b is a nonuniversal number and $T_{BKT}$ is the BKT transition temperature. The red line in Fig. 6 (f, g) presents the best fit to the above equation to the experimental data for the end members of the series i.e. $Ni_2P_2S_6/Fe_2P_2S_6$ with the obtained $T_{BKT}$ = 73.5/40K, and b = 0.756/0.45, respectively. We observe that the systems under considerations shows BKT physics. In particular, the temperature evolution of the correlation length fits well using BKT scaling. The extracted $T_{BKT}$ from the fits is below $T_N$ and is close to the transition temperature also reflected in the phonon renormalization.

## Conclusion

In summary, we performed a comprehensive temperature and doping dependent inelastic light scattering measurements (Raman study) on quasi two-dimensional magnetic van der Waals single crystals of $(Ni_xFe_{1-x})_2P_2S_6$. By focusing on the anomalies in the observed phonon modes we extracted long-range antiferromagnetic ordering temperature marked by the sharp changes in the phonons peak position, linewidths and intensities. Additionally, our observation of a broad magnetic continuum and its anomalous broad linewidth and coupling with the nearby phonon mode hints for a strong magnetostrictive coupling. From the two-magnon peak energy, the exchange interaction parameters is also estimated and found to vary by ~ 100 percent. Our results also evince the possible topological phase transition at a temperature much below the antiferromagnetic ordering of the spins, opening the possibility of identifying the topological ordering of spins using Raman scattering as a probe in quasi 2D magnetic systems.

**Acknowledgement:** NK acknowledge CSIR India for the fellowship. PK acknowledge support from IIT Mandi for the financial and experimental facilities. SA acknowledges financial support of DFG via Grant No AS 523/4-1. BB via SFB 1143 project-id 247310070), and Würzburg-Dresden Cluster of Excellence on Complexity and Topology in Quantum Matter–ct.qmat (EXC2147, project-id 390858490).

**Figure Captions:**

**Figure 1:** Raman spectrum of $(Ni_xFe_{1-x})_2P_2S_6$ collected at 4 K excited with 532 nm laser for (a) x=1, (b) x=0.7, (c) x=0.5, (d) x=0.3 and (e) x=0. The solid red lines are the total sum of Lorentzian fit to the experimental data and solid green lines are the individual fits of phonon modes. The observed phonon modes are labelled as P1-P10, S1-S4 and J1, J2. The observed broad peak (shaded part) is attributed to the two-magnon (2M) excitations. Spectral weight of 2M continue to decrease with increasing Fe percentage and vanishes completely for x=0. Inset (a) right side show peak P8 showing Fano asymmetry and inset (a) left side illustrate the raw Raman spectra of $Ni_2P_2S_6$ where 2M peak is shown.

**Figure 2:** (a) Temperature dependent frequency and FWHM of P1, P3, P4, P6, P7 and P9 phonon modes for $Ni_2P_2S_6$. (b) Temperature dependent frequency and FWHM of J1 and P2-P6 phonon modes for $Fe_2P_2S_6$. The blue and red dashed lines indicate the BKT ($T_{BKT}$), anti-ferromagnetic ($T_N$) transition temperature, respectively. The green shaded region shows the BKT region. Blue, magenta and red solid and semi-transparent lines are guide to the eye.

**Figure 3:** Evolution of the BKT (square) and Néel temperature ($T_N$) (sphere -Raman data, star-magnetic data) as a function Ni doping in $(Ni_xFe_{1-x})_2P_2S_6$ as extracted from our Raman measurements. PM represents paramagnetic phase above $T_N$ and orange colour region represent topological ordered phase below $T_{BKT}$.

**Figure 4:** (a) Two-magnon (2M) continuum (shaded region) for members of $(Ni_xFe_{1-x})_2P_2S_6$ at 4 K. For all the doping, intensity is plotted on the same scale clearly showing the reduction in 2M spectral weight with increasing Fe percentage. Temperature dependence of (b) Frequency, (c) FWHM and (d) Intensity of the 2M peak for different degree of Ni percentage. Dotted and semi-transparent (blue) solid lines are guide to the eye.

**Figure 5:** (a) Temperature evaluation of the Fano peak (P8 mode) plotted on the same intensity scale for all the temperature for $Ni_2P_2S_6$. (b) and (c) Temperature-dependence of the frequency and FWHM of the Fano peak, respectively. (d) Temperature dependence of the Fano asymmetry parameter $(1/|q|)$. Vertical blue and red dashed lines indicate the BKT ($T_{BKT}$) and antiferromagnetic ($T_N$) transition temperature. The green shaded region shows the BKT region. Semi-transparent (blue) solid lines are guide to the eye.

**Figure 6: (a-e)** Temperature dependence of the dynamic Raman susceptibility ($\chi^{dyn}$) for $(Ni_xFe_{1-x})_2P_2S_6$, where x = 0 to 1, extracted using the Kramer-Kronig relation as described in the text, integrating up to ~ 56 $cm^{-1}$. Semi-transparent (blue) solid lines are guide to the eye. (f-g) Shows temperature dependence of the spin-spin correlation length [$\xi(T)$] for $Ni_2P_2S_6$ and $Fe_2P_2S_6$ extracted from the Raman conductivity above ~ 150K and 120K, respectively. Inset shows $\xi(T)$ in the full temperature

window i.e. from 4 to 330K. Red solid lines are the fit to $\xi(T)$ by BKT exponential scaling law expressed as $\xi(T) \sim \exp(b\sqrt{T_{BKT} / T - T_{BKT}})$.

**Figure1:**

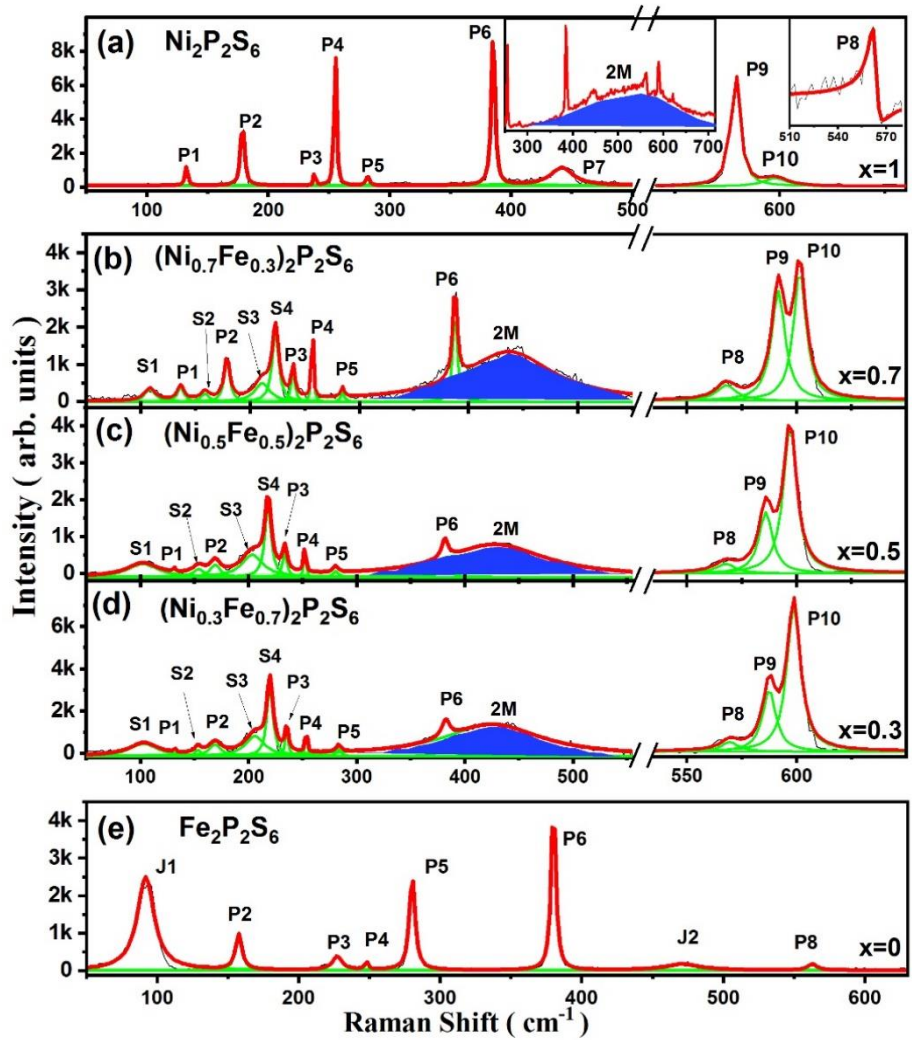

**Figure 2:**

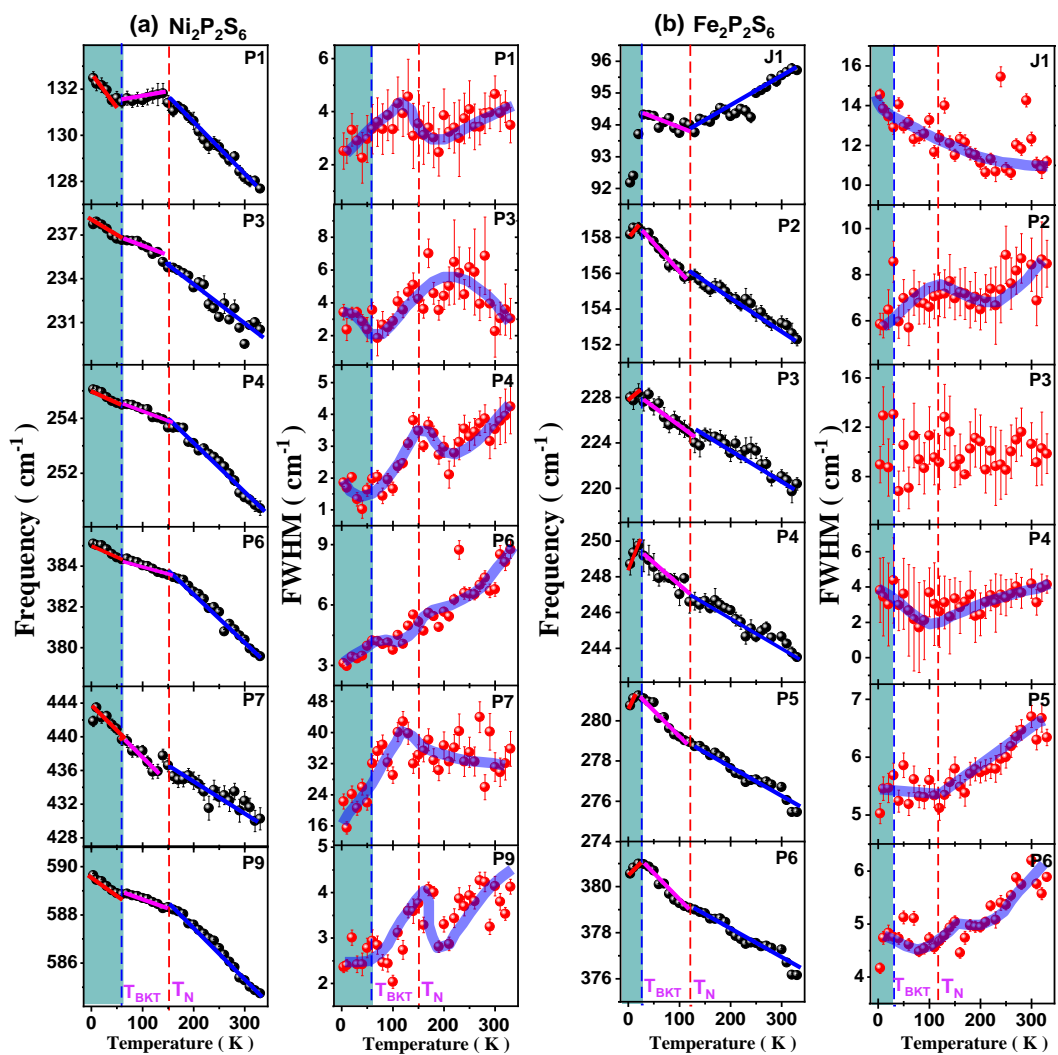

**Figure 3:**

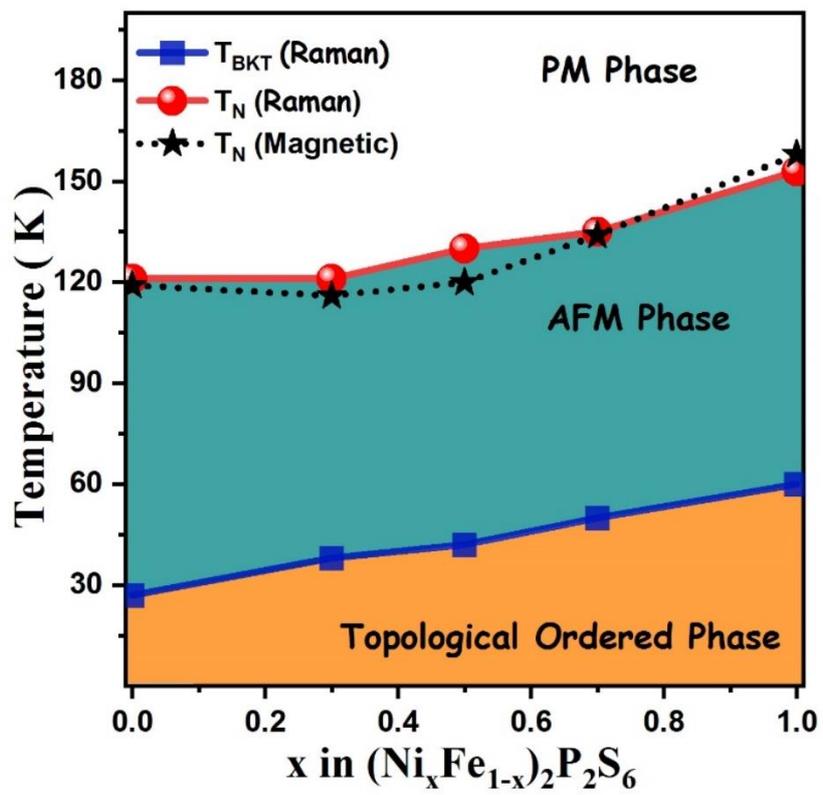

**Figure 4:**

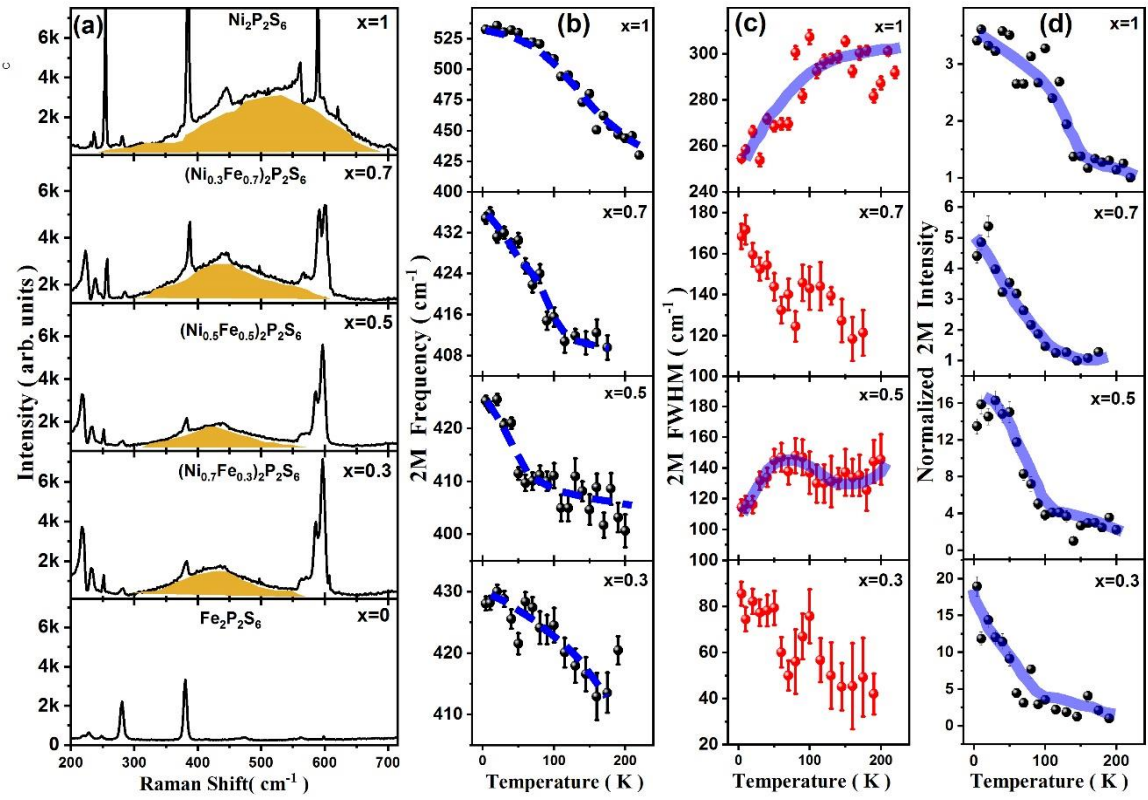

**Figure 5:**

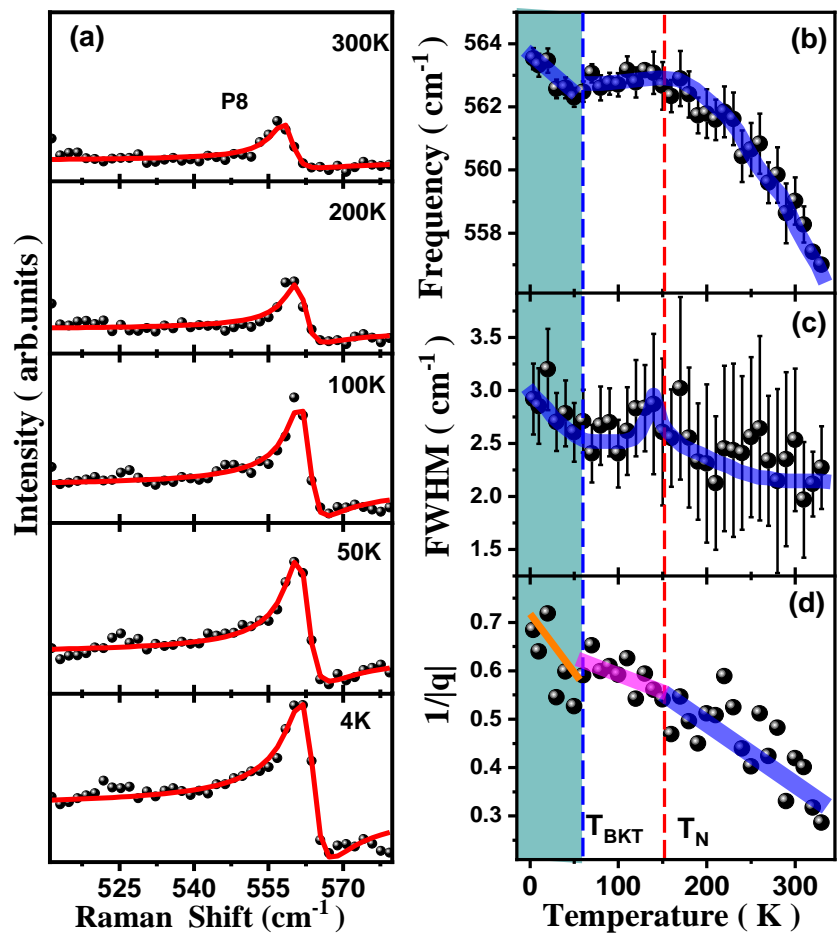

**Figure 6:**

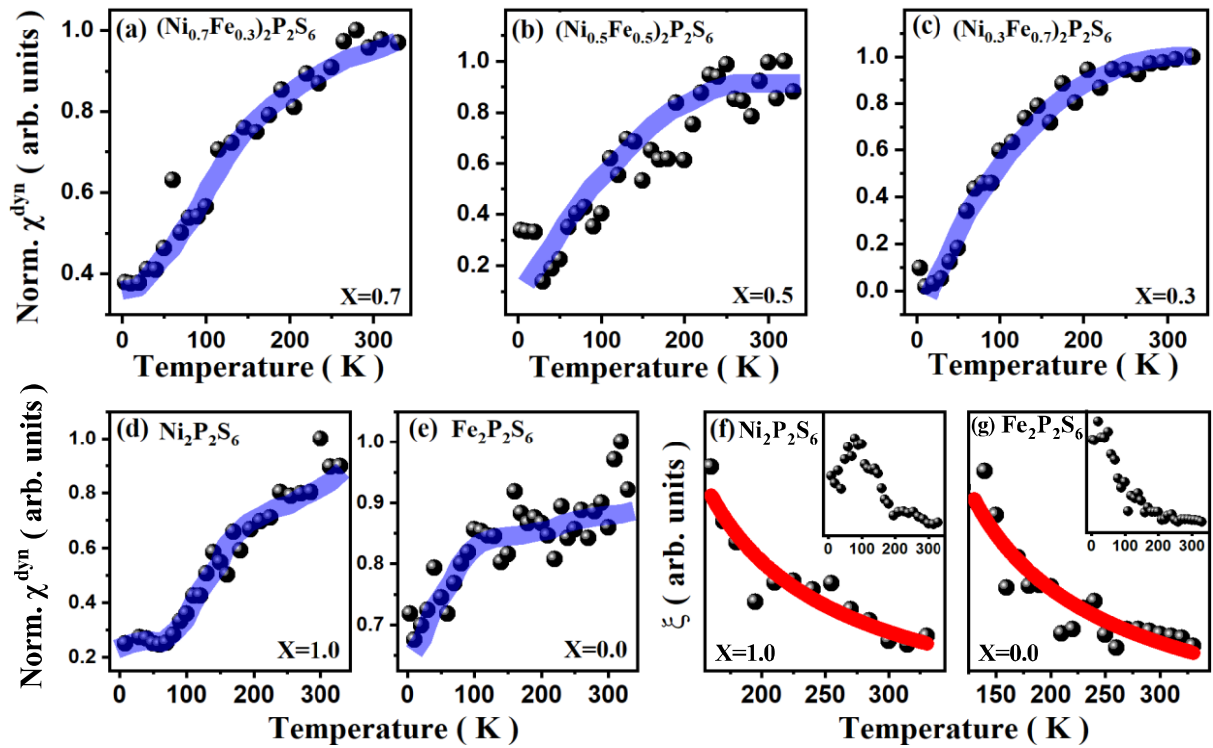

# Interplay of topology and antiferromagnetic order in two-dimensional van der Waals crystals of $(Ni_xFe_{1-x})_2P_2S_6$

N. Khan[1,*], D. Kumar[1], V. Kumar[1], Y. Shemerliuk[2], S. Selter[2], B. Büchner[2,4], K. Pal[3], S. Aswartham[2] and Pradeep Kumar[1,†]

[1] School of Physical Sciences, Indian Institute of Technology Mandi, Mandi-175005, India

[2] Leibniz-Institute for Solid-state and Materials Research, IFW-Dresden, 01069 Dresden, Germany

[3] Department of Physics, Indian Institute of Technology Kanpur, Kanpur- 208016, India

[4] Institute of Solid State and Materials Physics and Würzburg-Dresden Cluster of Excellence ct.qmat, Technische Universität Dresden, 01062 Dresden, Germany

[*]nasarukhan736@gmail.com
[†]pkumar@iitmandi.ac.in

## Section-I: Anharmonic phonon-phonon interaction model

To understand the normal temperature dependency of peak frequency and FWHM of the phonon modes i.e., peak frequency decreases and FWHM increases with increasing temperature, we fitted the phonon mode frequency from Néel temperature to 330 K with an anharmonic phonon-phonon interaction model which is called three and four phonon model. In the anharmonic phonon model fitting, we have taken into account both cubic anharmonicity i.e., three phonon interaction in which an optical phonon decays into two acoustic phonons and quadratic anharmonicity i.e., four phonon interaction in which an optical phonon decays into three acoustic phonons. According to this model phonon frequency and linewidth are given by the following expressions [1]

$$\omega(T) = \omega_0 + C\left[1 + \frac{2}{e^x - 1}\right] + D\left[1 + \frac{3}{(e^y - 1)} + \frac{3}{(e^y - 1)^2}\right] \qquad (1)$$

$$\Gamma(T) = \Gamma_0 + A[1 + \frac{2}{e^x - 1}] + B[1 + \frac{3}{(e^y - 1)} + \frac{3}{(e^y - 1)^2}] \tag{2}$$

Where $\omega_0$ and $\Gamma_0$ are the phonon frequency and linewidth at absolute zero temperature respectively, and C, D are self-energy parameter which are constant and $x = \frac{\hbar\omega_0}{2k_B T}$, $y = \frac{\hbar\omega_0}{3k_B T}$. Figures S2 and S3 Solid red line represents anharmonic model frequency fits and solid blue red lines indicates the anharmonic model FWHM fits, from $T_N$ to 330 K for the entire series. Temperature variation of phonon mode frequency in this range is well explained by the cubic and quadratic anharmonicity as shown in the figures S2 and S3, however some weak modes could not be fitted as the data is scattered. Linewidth of some of the phonon modes is well explained by cubic and quadratic anharmonicity. The values of C, D, A, B obtained by fitting are given in the Table S4 and Table S5. It can be noted from the values given in table that three phonon interaction is dominant in this temperature region. Below $T_N$ deviation from anharmonic interaction model fitting in phonon frequency can be observed clearly. Below $T_N$ magnetic degree of freedom comes into existence, hence coupling between lattice and magnetic degree of freedom needs to be taken into account and simple phonon anharmonic model is not sufficient.

**Section-II: Experimental and computational details**
**Experimental details:**

Temperature dependent Raman measurements of single crystals of $(Ni_xFe_{1-x})_2P_2S_6$ series were done using Horiba LabRAM HR Evolution Raman spectrometer in the backscattering geometrical configuration and in the temperature range between 4 K to 330 K with the temperature accuracy $\pm 0.1$ K. The samples were illuminated with a laser of 532 nm wavelength using a 50X long working distance objective for focusing as well as collecting the light scattered from the sample. To avoid any local heating effect on the sample surface (Peak parameters are affected by heating effect) the laser power was taken very low, $< 1$ mW. The light scattered from the sample was detected by a Peltier cooled Charge Coupled Device using the grating 600 and 1800 groove per mm. A closed cyclic refrigerator (Montana instrument) was used to get the very low- temperature.

In this work all studies were done on high-quality single crystals of $(Ni_xFe_{1-x})_2P_2S_6$ which were grown by the chemical vapor transport technique with iodine as the transport agent. All details of the growth and physical characterization of the crystals studied are described in ref. [2].

**Computational details:**

The first-principles calculations based on density functional theory (DFT) were performed using the Vienna Ab-initio Simulation Package (VASP) [3] utilizing the PBE [4] generalized gradient approximation (GGA) to the exchange-correlation energy functional. We used the projector augmented wave (PAW) [5-6] potentials for Fe ($3d^6 4s^2$), Ni ($3d^8 4s^2$), S ($3s^2 5p^4$) and P ($3s^2 3p^3$). We included the van der Waals interactions in all calculations using the DFT-D3 method of Grimme [7]. We considered non-magnetic configurations for optimization and the calculations of phonons for $Ni_2P_2S_6$. The optimized lattice parameters of $Ni_2P_2S_6$ are in good agreement with the experimental values (experiment: $a = 5.811$ Å, $b = 10.089$ Å, $c = 6.628$ Å; theory: $a = 5.765$ Å, $b = 10.005$ Å, $c = 6.491$ Å). For $Fe_2P_2S_6$, the optimized lattice parameters show good agreement when we considered a magnetic unit cell (antiferromagnetic configuration). The optimized lattice parameters for $Fe_2P_2S_6$ are $a = 5.930$ Å, $b = 10.310$ Å, $c = 6.720$ Å that agree well with the experimental values $a = 5.881$ Å, $b = 10.346$ Å, $c = 6.642$ Å. The phonon frequencies for $Fe_2P_2S_6$ were calculated afterwards utilizing these optimized lattice parameters. We used a kinetic energy cut-off of 520 eV for DFT calculations and a k-point mesh of 10 x 10 x 10 for the primitive unit cell and scaled properly for the supercells. To calculate the phonon dispersion, we used supercell based finite-difference method. The phonon dispersions were calculated using $2\times2\times2$ supercell (containing 80 atoms) of the primitive unit cell (having 10 atoms) using Phonopy [8]. While $Ni_2P_2S_6$ does not exhibit any imaginary phonon frequencies (Fig S10a), $Fe_2P_2S_6$ shows some unstable phonon modes. To stabilize the unstable phonon modes of $Fe_2P_2S_6$, we freeze the atoms in the unit cell of $Fe_2P_2S_6$ with the eigenvector of the most unstable phonon mode at $\Gamma$ point and optimized the unit cell again. This optimized unit cell of $Fe_2P_2S_6$ lowered its energy and symmetry compared to the parent high-symmetry unit cell. Phonon calculations were performed taking this unit cell which stabilizes the $\Gamma$ point phonon frequencies (Fig. S10c).

The primitive unit cell of $Ni_2P_2S_6$ and $Fe_2P_2S_6$ contain 10 atoms. Hence, there will be a total number of 10x3 = 30 phonon modes. Out of these, 3 are acoustic modes with zero frequency at $\Gamma$ point and 27 are optical modes. Since only $\Gamma$ point phonons are primarily relevant for the experimentally observed Raman modes, we provide all the $\Gamma$ point optical phonon frequencies for $Ni_2P_2S_6$ (107, 112, 119, 138, 141, 145, 158, 158, 192, 197, 198, 215, 219, 224, 225, 228, 229, 246, 247, 251, 350, 410, 504, 526, 527, 532, 534 $cm^{-1}$) and $Fe_2P_2S_6$ (107, 150, 162, 170, 191, 197, 215, 215, 219, 222, 226, 228, 234, 236, 269, 271, 277, 292, 300, 318, 325, 337, 340, 366, 483, 483, 492 $cm^{-1}$). To compare the calculated phonon frequencies with the experimental values, we choose the frequencies of $Ni_2P_2S_6$ for which the optimized unit cell retains the space group symmetry of the experimental crystal structure and hence, assignment of symmetry labels become much convenient. The relevant phonon frequencies are given in Table S1 and their eigenvector visualizations are given in Fig. S11.

**Section-III: Origin of two-magnon**

Two-magnon (2M) scattering have similar origin as the second order scattering by the phonons. Considering the spin-orbit mechanism (Elliot-Loudon process) for magnetic scattering [9-10] and extending it to the case of second-order, the scattering cross-section is fourth order in the spin-orbit coupling strength and is generally several orders of magnitude weaker than the 1st order magnetic cross-section and as a result 2M would be very weak and hardly observable. However, experimentally it is found that intensity of 2M is comparable or even more than the one-magnon. These experimental observations on variety of systems points that $2^{nd}$ order magnetic scattering is not just a simple extension of $1^{st}$ order magnetic scattering (which have its origin in the spin-orbit mechanism) rather have an entirely new phenomena which occurs only in case of $2^{nd}$ order. Such a mechanism has been proposed and is well understood, known as the Exchange Scattering Mechanism (ESM) [11-13]. Pictorial representation of ESM is shown in Fig. S0(a). There are basically three steps involved i.e. (i) absorption of incident photon, (ii) emission of scattered photon on $i^{th}$ ion and absorption of scattered photon by $j^{th}$ ion, (iii) emission of scattered photon on $j^{th}$ ion. This process leads to excitation of both the ions from their ground state ($\left| GS \right\rangle$) to their first excited spin states. The main feature of this ESM is that spin deviations are created at a pair of exchange coupled magnetic sites on opposite sublattice (owing to the constraint that z-component of the spins is conserved i.e., $\Delta S^Z = 0$).

For our current study, surprisingly, we observed strong two-magnon spectral weight only for the case of $Ni_2P_2S_6$ and Ni doped samples suggesting dominant role of ESM. On the other hand, for the case of $Fe_2P_2S_6$, we do not see any two-magnon signal hinting minimal or no role of ESM in this particular case. We have tried to understand this anomaly invoking underlying nature of insulator these materials belongs to, exchange interaction between transition metals via surrounding ligands (sulphur) and the resonance involving phonon modes associated with the $(P_2S_6)$ cage. It has been established that $Fe_2P_2S_6$ is a Mott-Hubbard type insulator with $\Delta_{pd} > U_{dd}$, whereas $Ni_2P_2S_6$ is a charge transfer type insulator (also termed as self-doped negative charge transfer insulator) with $\Delta_{pd} < U_{dd}$; where $\Delta_{pd}$ is the energy required for charge transfer between $d$-state of Ni/Fe and $p$-states of the ligand (sulphur), and $U_{dd}$ is the $d$-$d$ coulomb interaction energy [14-16]. Typical values of these parameters for $Fe_2P_2S_6$/ $Ni_2P_2S_6$ are: $\Delta_{pd} \sim$ 6eV/-1eV & $U_{dd} \sim$ 3eV/5eV. For the case of $Ni_2P_2S_6$ a direct exchange of electron from one Ni atom to the next Ni atom is not favoured due to energy constrained i.e., $\Delta_{pd(-1eV)} < U_{dd(5eV)}$. However, an ESM, via mid atoms i.e., Sulphur ($Su$), facilitated by transfer of an electron from $Su$ to $Ni$ and vice versa (owing to self-doped negative charge transfer insulator nature) may be more effective as Ni atom is surrounded by six $Su$ atoms [see the dash yellow colour circle in Fig. S0 (b)]. We note that such a super-exchange process between Ni atoms mediated by Su atom, in addition to the direct exchange between Ni atoms, has also been treated theoretically in a recent report [17]; and it was shown that $J_1^s$ (super-exchange between Ni atoms mediated via sulphur) is order of magnitude bigger than $J_1^D$ ( direct exchange between Ni atoms). Also $J_3^s$ is much larger than $J_3^D$ for the case of $Ni_2P_2S_6$.

On the other hand, $Fe_2P_2S_6$ is a Mott-Hubbard type insulator with $\Delta_{pd(6eV)} > U_{dd(3eV)}$ and therefore probability of super-exchange of electron via Su atom is minimal. This will significantly reduce the ESM efficiency, as only direct exchange route is possible, responsible for the two-magnon which requires exchange of electron to Fe atoms on opposite sublattice (i.e. only $J_2$ and $J_3$ play a role here). It is to be noted that direct overlap is negligible for the 2nd and 3rd nearest neighbour Ni/Fe atoms, since $d$-orbitals are localized in nature and the overlap between $d$-orbitals decays as $\sim 1/r^5$ [18]. See Fig. S0(b), $J_3$ (represents the exchange interaction strength between Fe atom with its 3rd nearest neighbour) distance is maximum so in case of $Fe_2P_2S_6$ its magnitude will be significantly reduced as it is possible only via direct exchange (see dotted blue colour line). For the case of $Ni_2P_2S_6$ the possible paths are via Sulphur atoms as well (see dotted red colour curve) and hence $J_3$ magnitude is expected to be much larger. A large value of $J_3$ in case of $Ni_2P_2S_6$ comes mainly from the super-exchange process ($J_3^s$) mediated by the sulphur atoms. Since super-exchange via sulphur in case of $Fe_2P_2S_6$ is not possible or is minimal, so main contribution to $J_3$ is via direct exchange ($J_3^D$) process hence very small. Recently, it has been suggested in a theoretical report for the case of two quantum dots that super-exchange processes enhance strongly as well range of spin-spin exchange increases significantly via mediation by a third quantum dots in between [19]. It has also been shown that super-exchange enhances dramatically if the angle between two quantum dots and the third dot in between is more than ninety degrees. For the case of $Ni_2P_2S_6$ the angle between two 3rd nearest neighbour Ni atoms and the sulphur ions in between is more than ninety degrees [17]. Based on the analogy mentioned above, we expect something similar for the case of $Mn_2P_2S_6$, as it is a Mott-Hubbard type insulator, hence absence of two-magnon. Indeed in case $Mn_2P_2S_6$ no two-magnon has been reported [20].

We believe that such as a strong ESM between Ni atoms involving sulphur atoms may also be reflected via phonon modes. The honeycomb lattice formed in the *ab*-plane by the transition metal atoms (Ni/Fe) encircled the $P_2S_6$ cage [see the dashed grey colour circle in Fig. S0(b)] and the phonon modes associated with this cage are the high energy ones with frequency > 250 $cm^{-1}$. We note that for the case of pure Ni or Ni doped samples we observe very strong high energy modes at ~ 600 $cm^{-1}$. The observation of strong phonon modes suggests a robust coupling between phonons associated with this cage and the charge transfer between Ni and Sulphur atoms or a resonance may be happening with the process of charge transfer and this resonance may give rise to these strong phonon modes. However, we do not see such phonon modes near ~ 600 $cm^{-1}$ in case of $Fe_2P_2S_6$ suggesting minimum or no coupling between phonons and charge transfer process, or no charge transfer is possible hence no resonance and as a result no observation of these high energy phonon modes.

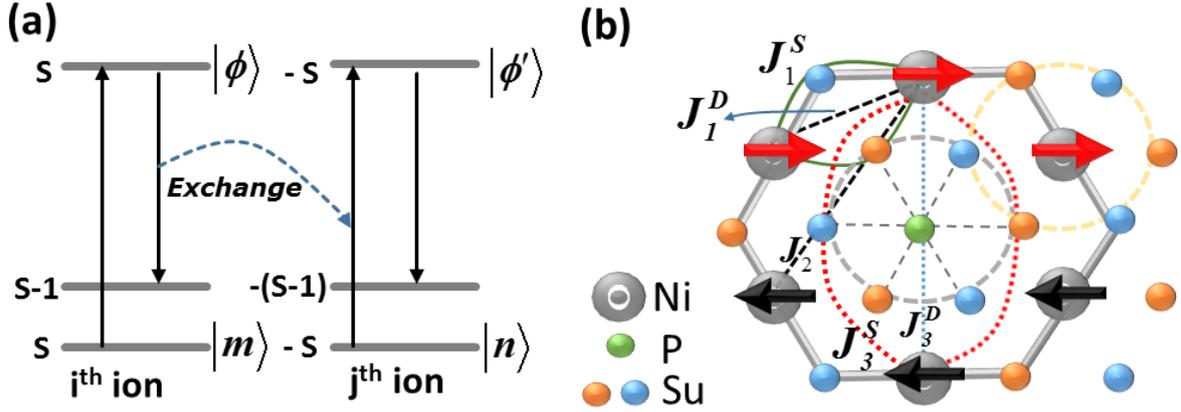

**Figure S0: (a)** Successive transition of an Exchange Scattering Mechanism for the case of two magnetic ions ($i^{th}$ and $j^{th}$) on the opposite magnetic sublattices (since the Hamiltionian conserve the total $z$-component of the electron spins ($\Delta S^Z = 0$); $\hat{H} = e(\hat{E}_i + \hat{E}_S).(\overline{r_1} + \overline{r_2}) + \dfrac{e^2}{4\pi\varepsilon_0 |\overline{r_1} - \overline{r_2}|}$; $\hat{E}_i / \hat{E}_S$ represent the incident/ scattered field). $|m\rangle$ and $|n\rangle$ are the ground state of $i^{th}$ and $j^{th}$ atom, respectively; and $S$ is the spin state. (b) Honeycomb lattice formed by the transition metal atoms in the *ab*-plane. Blue/orange colour sulphur atoms are above/below the plane. Metal atom (Ni/Fe) is surrounded by the six sulphur atoms forming an octahedra (see dashed yellow colour circle). $J_1^D / J_1^S$ and $J_3^D / J_3^S$ represents the direct/super exchange interaction between the first and third nearest neighbour Ni/Fe atoms, respectively. $J_2$ represents the direct exchange interaction between first and $2^{nd}$ nearest neighbour Ni/Fe atoms. Dotted blue/red colour line shows the direct/super-exchange interaction between $3^{rd}$ nearest neighbour Ni atoms. Dashed grey colour circle shows the $P_2S_6$ cage, $P_2$ dimer (green colour) is at the centre.

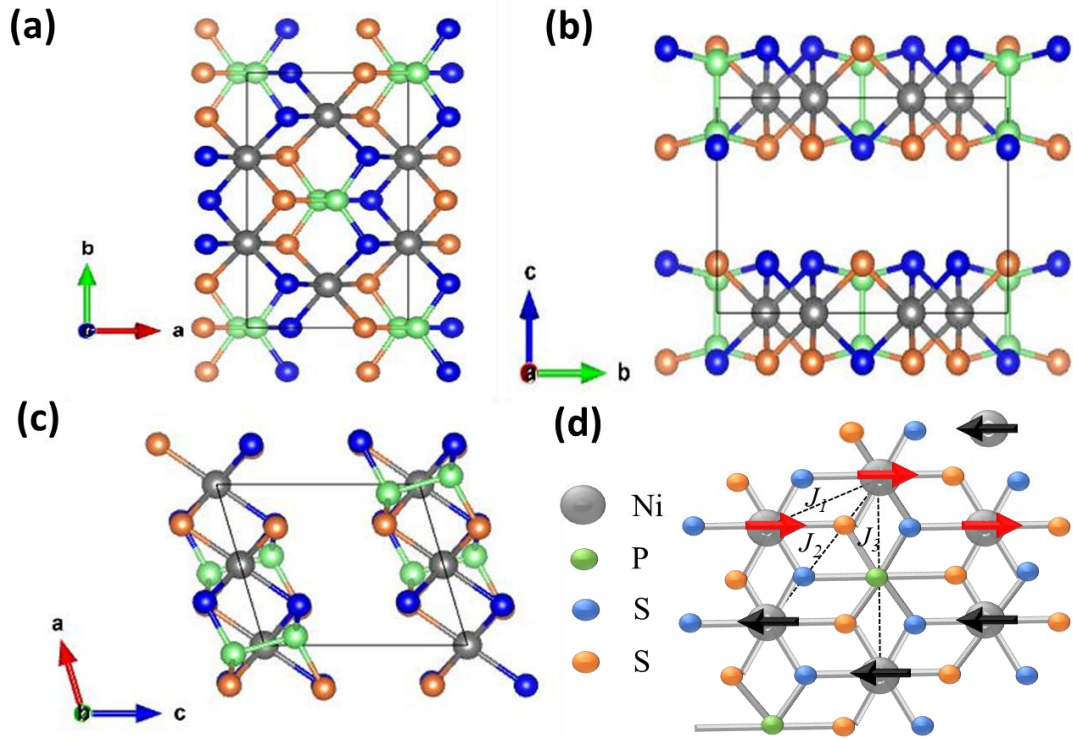

**Figure S1:** The crystal structure of TM$_2$P$_2$S$_6$ (TM = Ni, Fe) kind of materials shown in (a) *ab* plane (b) *bc* plane (c) *ac* plane. The representation was prepared using VESTA software. (d) Shows the spin structure of Ni$_2$P$_2$S$_6$ below $T_N$ in the *ab*-plane. Red/black colour arrow on Ni atoms represents spins.

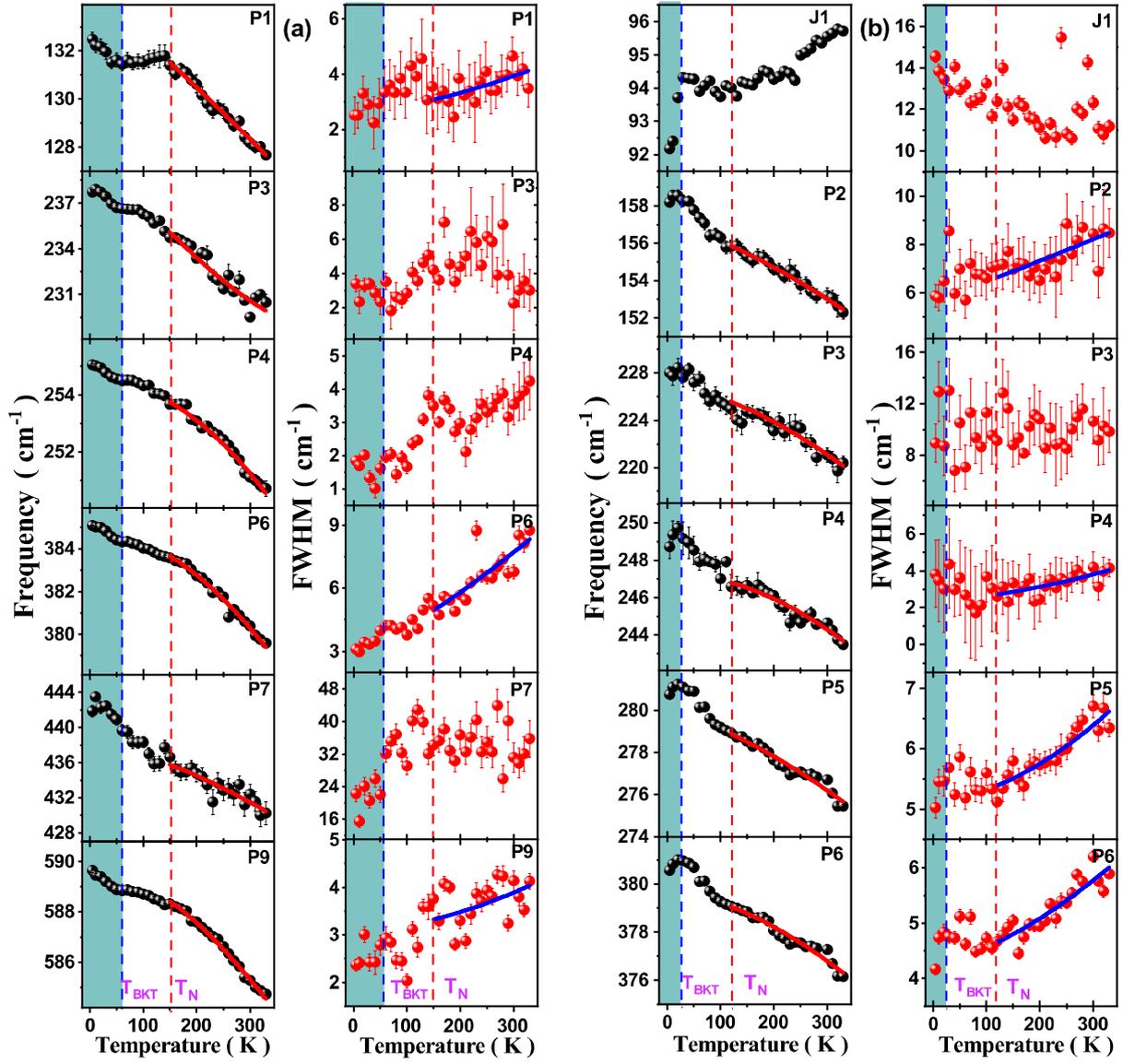

**Figure S2:** (a) Temperature dependent frequency and FWHM of P1, P3, P4, P6, P7 and P9 phonon modes for $Ni_2P_2S_6$. (b) Temperature dependent frequency and FWHM of J1 and P2-P6 phonon modes for $Fe_2P_2S_6$. The blue and red dashed lines indicate the BKT ($T_{BKT}$), Anti-ferromagnetic ($T_N$) transition temperature, respectively. The green shaded region shows the BKT region. The red and blue solid lines indicate the anharmonic model fit for the frequency and FWHM, respectively, above $T_N$.

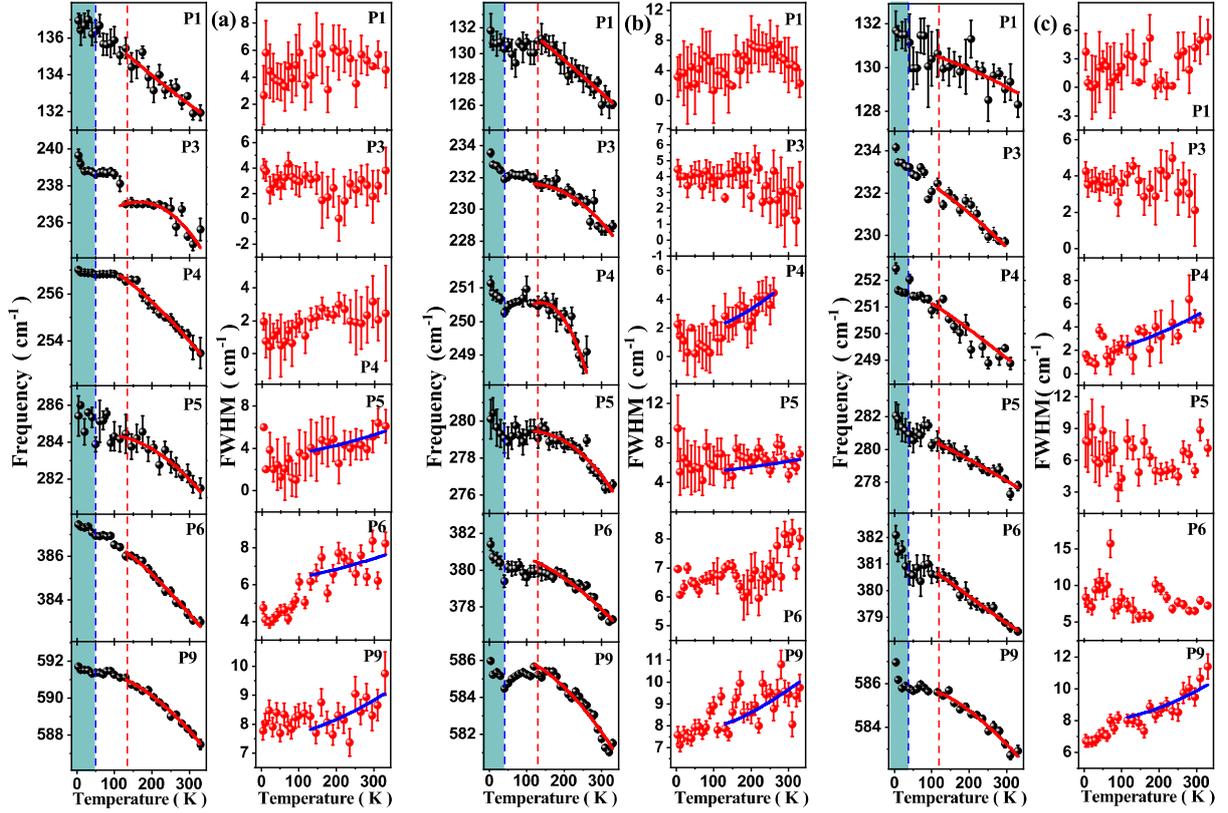

**Figure S3:** Temperature dependent frequency and FWHM of P1, P3, P4, P5, P6 and P9 phonon modes of $(Ni_xFe_{1-x})_2P_2S_6$ for (a) x = 0.7, (b) x = 0.5 and (c) x = 0.3, respectively. The blue and red dashed lines indicate the BKT ($T_{BKT}$), Anti-ferromagnetic ($T_N$) transition temperature, respectively. The green shaded region shows the BKT region. The red and blue solid lines indicate the anharmonic model fit for the frequency and FWHM respectively.

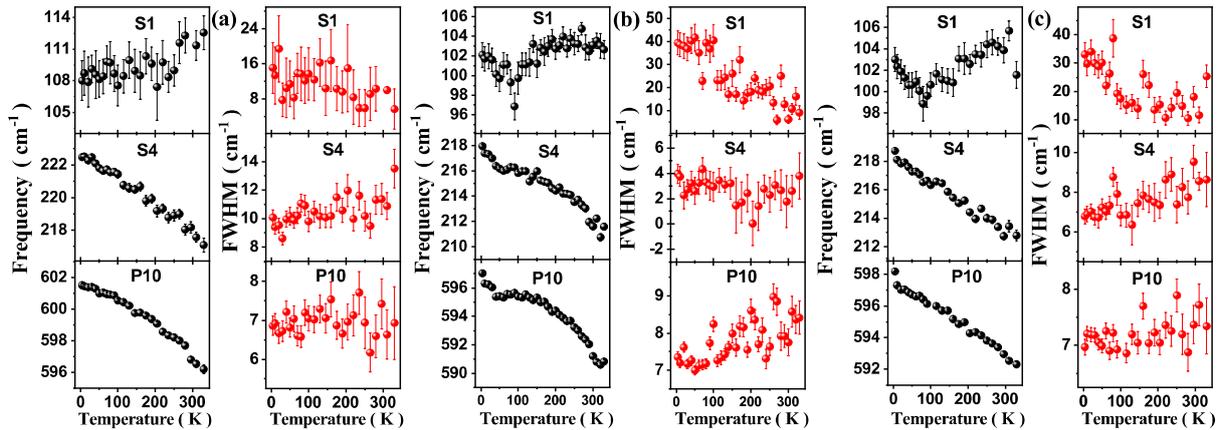

**Figure S4:** Temperature dependent frequency and FWHM of S1, S4 and P10 phonon modes of $(Ni_xFe_{1-x})_2P_2S_6$ for (a) x = 0.7, (b) x = 0.5 and (c) x = 0.3, respectively.

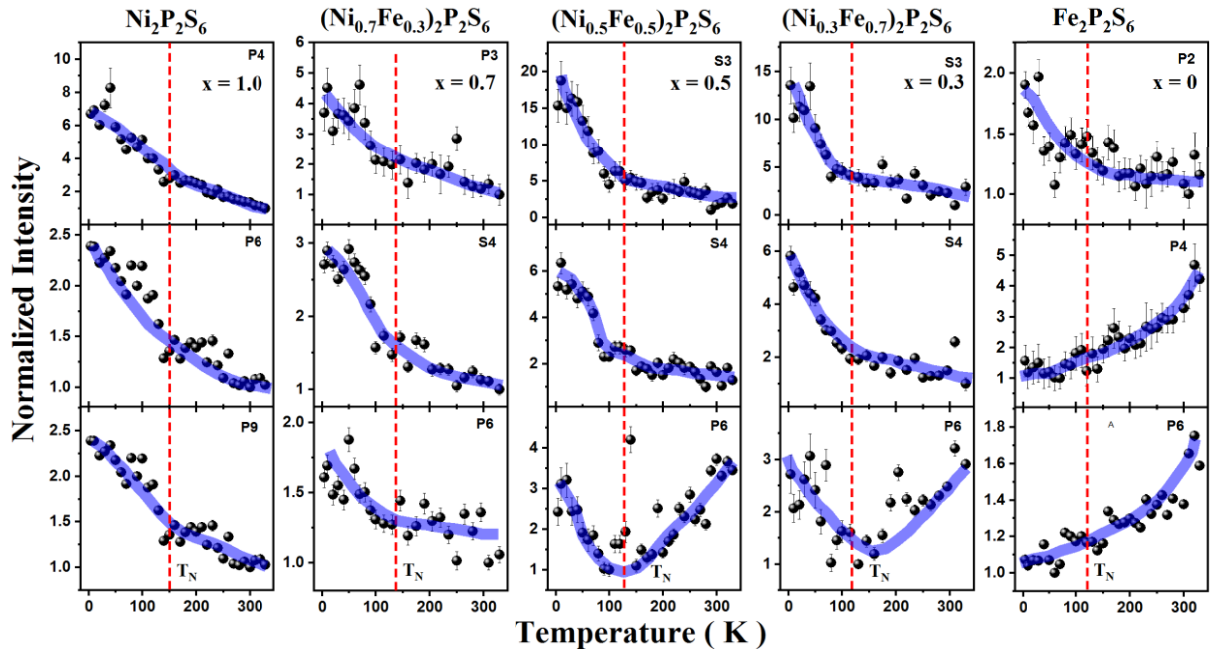

**Figure S5:** Temperature dependent intensity evolution of some of the prominent modes for all the members of $(Ni_xFe_{1-x})_2P_2S_6$.

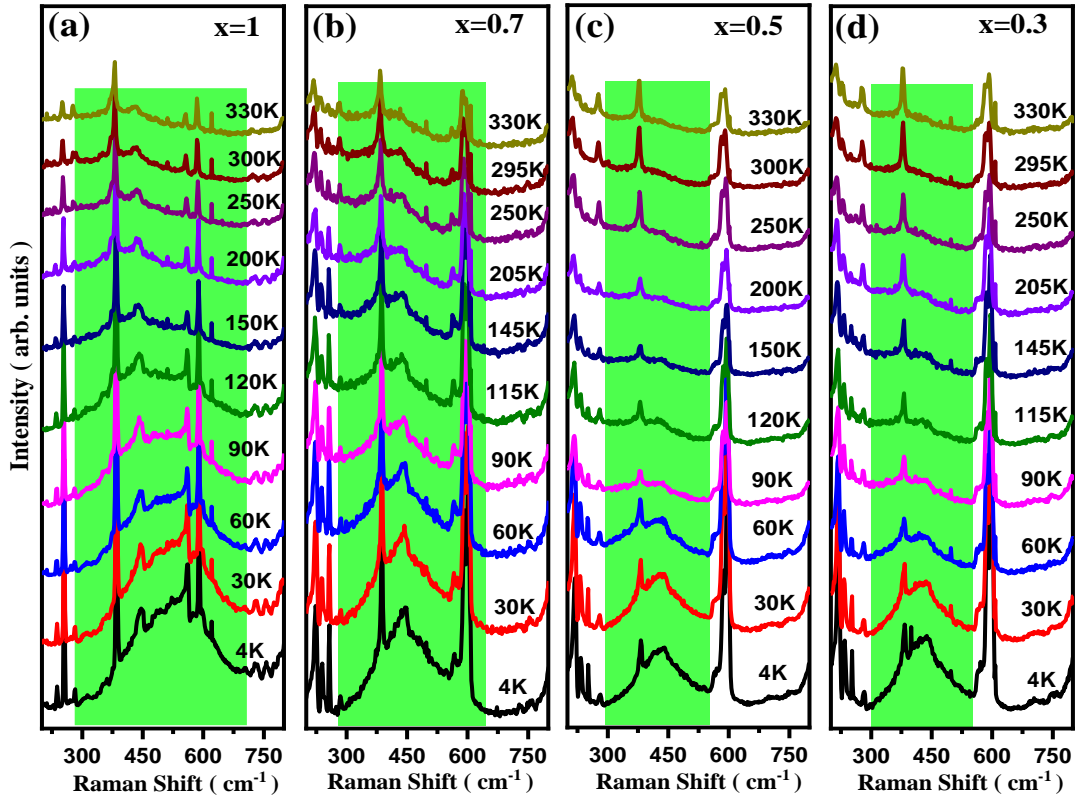

**Figure S6:** Temperature evaluation of the two-magnon signal (shaded region) of $Ni_xFe_{1-x}PS_3$ for (a) x=1, (b) x=0.7, (c) x=0.5, (d) x=0.3, respectively.

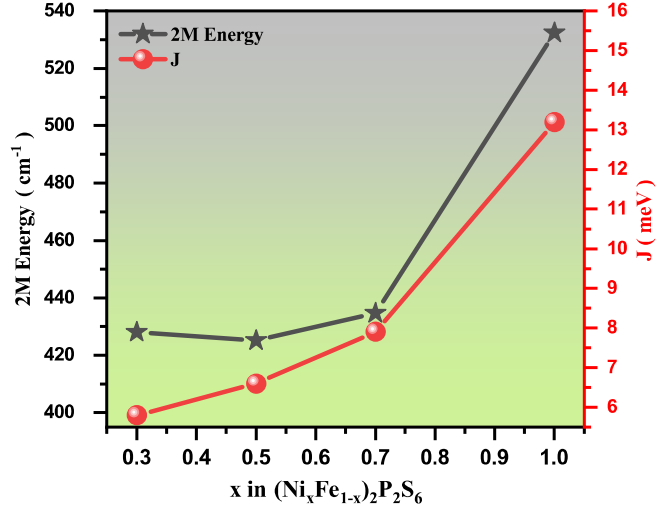

**Figure S7:** Two-magnon energy and nearest neighbour exchange parameter ( J ) as a function of increasing Ni doping for $(Ni_xFe_{1-x})_2P_2S_6$ series, where x is degree of Ni substitution.

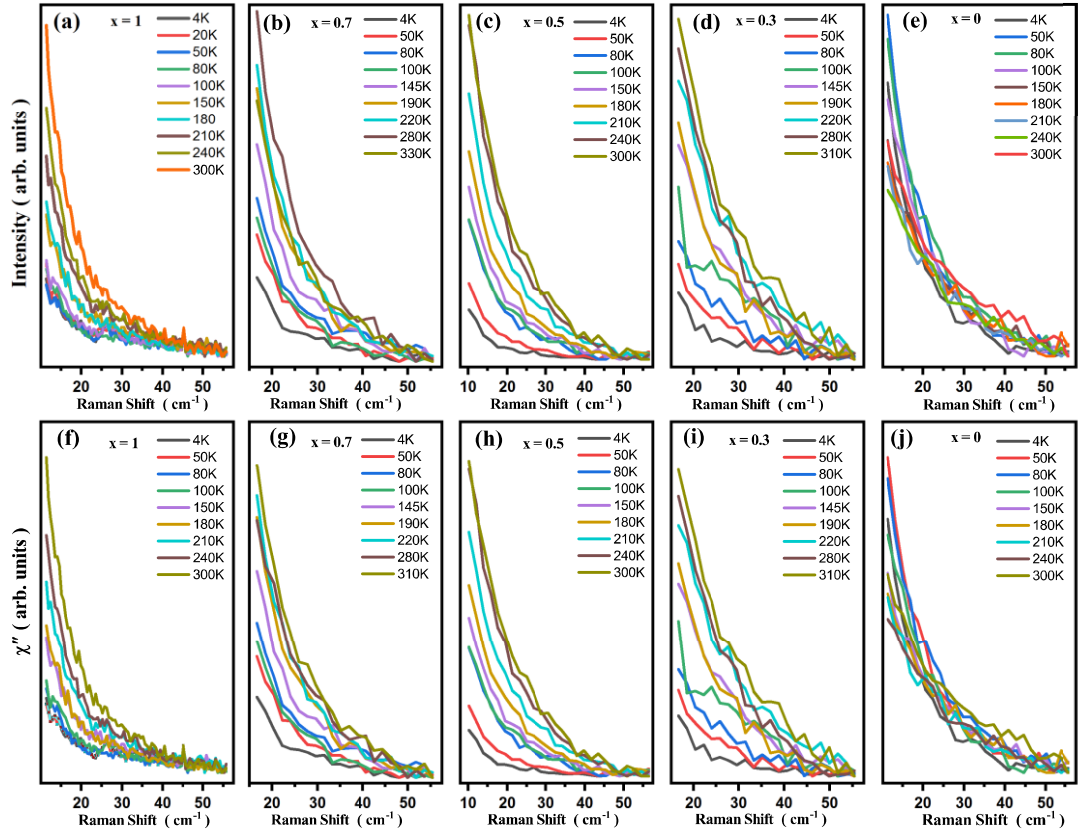

**Figure S8:** Temperature dependence of the quasi-elastic scattering (QES) signal in the frequency range of 10- 56 $cm^{-1}$ for $(Ni_xFe_{1-x})_2P_2S_6$ of **(a)** x = 1, **(b)** x = 0.7, **(c)** x = 0.5, **(d)** x = 0.3 and **(e)** x = 0, respectively. **(f-e)** Temperature dependence of the Raman response $\chi''(\omega)$ in the frequency range of 10- 56 $cm^{-1}$. QES is significantly suppressed below $T_N$, reflecting the setting up of the long-range magnetic ordering, as the spin fluctuations are expected to suppress below the transtion temperature.

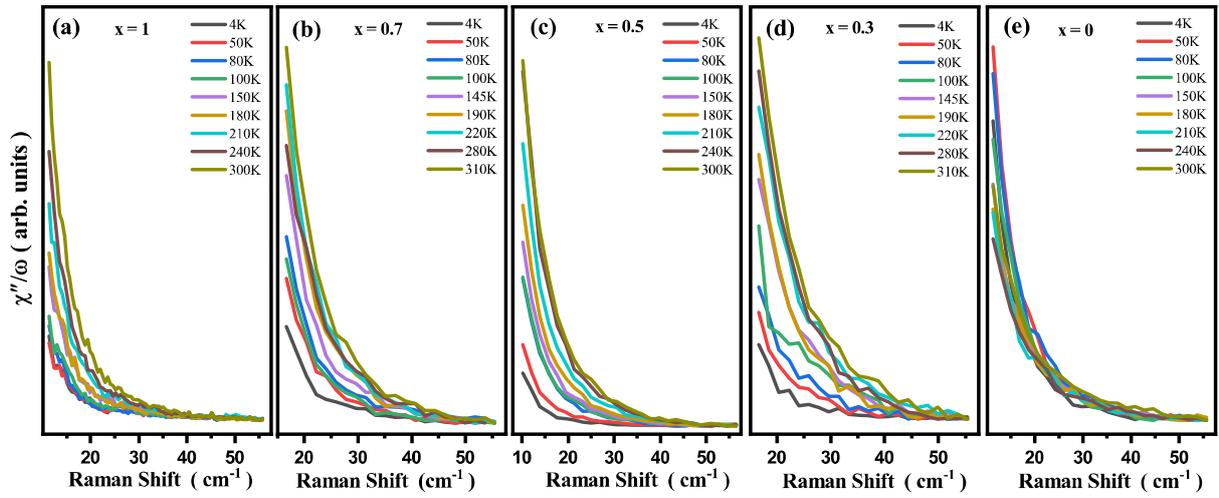

**Figure S9:** Temperature dependence of the Raman conductivity ($\chi''/\omega$) of (Ni$_x$Fe$_{1-x}$)$_2$P$_2$S$_6$ for (a) x = 1, (b) x = 0.7, (c) x = 0.5, (d) x = 0.3 and (e) x = 0, respectively.

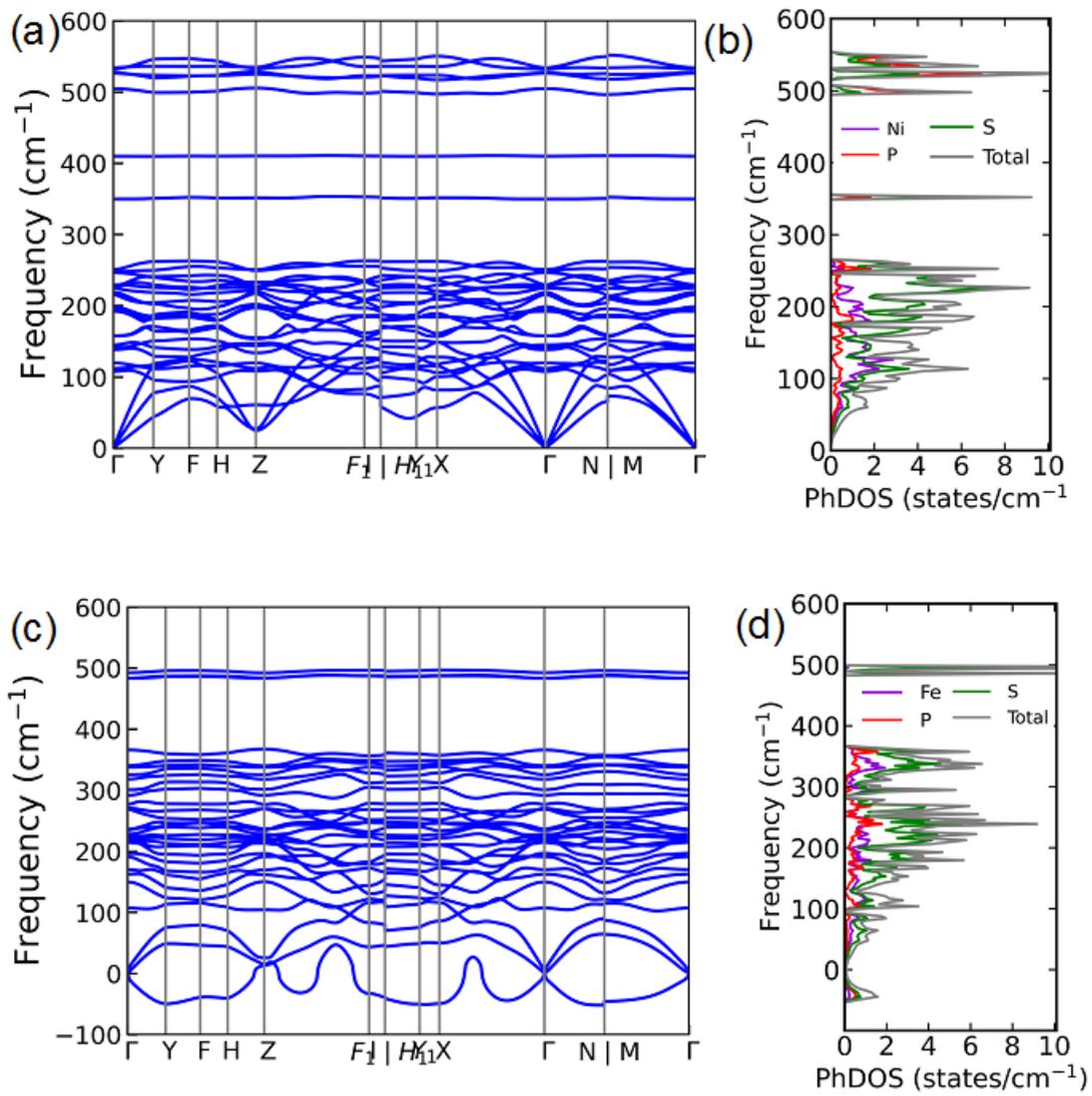

**Figure S10:** Calculated phonon dispersions and phonon density of states of $Ni_2P_2S_6$ (a, b) and $Fe_2P_2S_6$ (c,d).

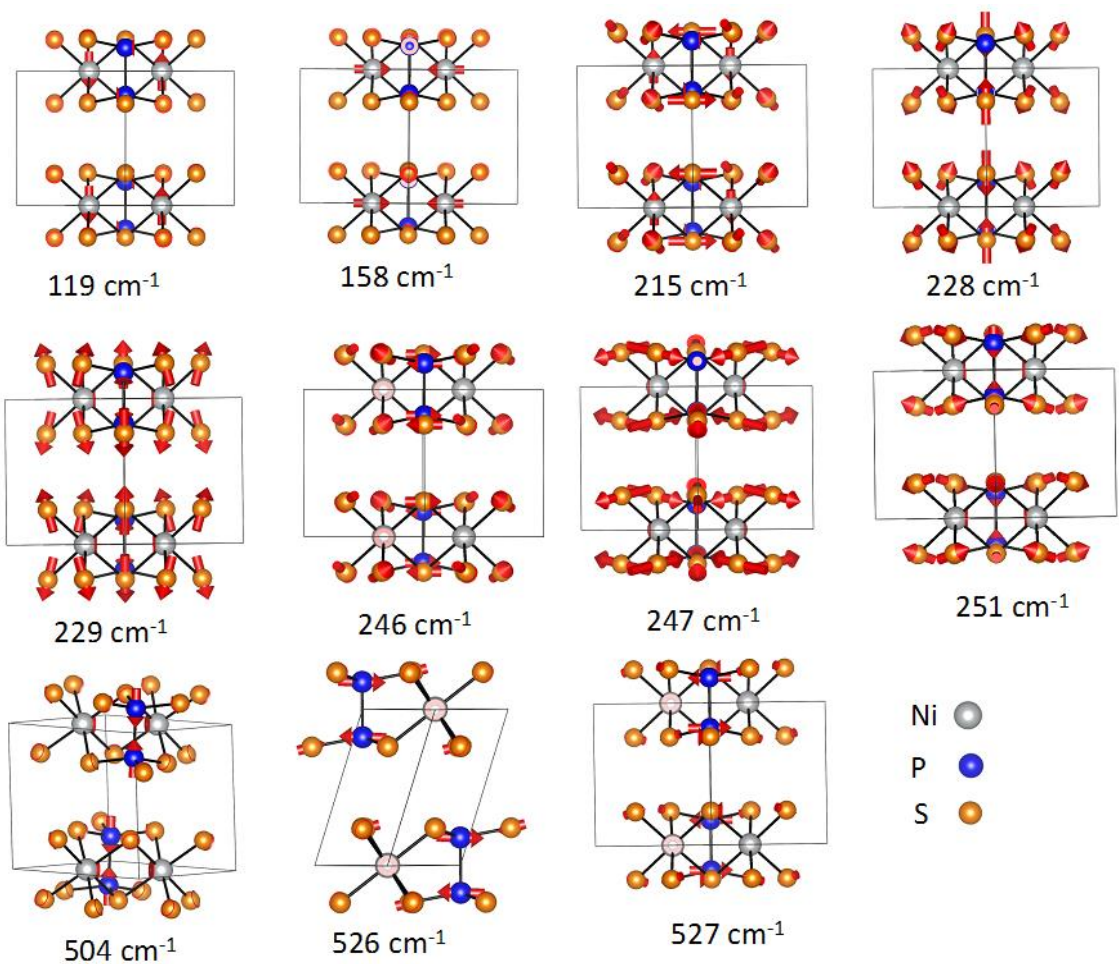

**Figure S11:** Eigen vectors for the relevant Γ point phonon modes of Ni$_2$P$_2$S$_6$, that are assigned to the experimentally observed Raman modes. Grey, blue, and orange spheres represent Ni, P and S atoms, respectively. Red arrows on the atoms represent the direction of motion and the size of the arrows represent the relative magnitude of vibration.

**Table-S1**: List of the experimentally observed modes (in cm$^{-1}$) along with their symmetry assignment and frequency at 4 K for (Ni$_x$Fe$_{1-x}$)$_2$P$_2$S$_6$ series. Mode assignment has been done on the basis of our first-principles calculations of phonon frequencies at Γ point for Ni$_2$P$_2$S$_6$. The calculated phonon frequencies are denoted by $\omega_{DFT}$.

| Mode | Frequency ($\omega$) at 4K | | | | | Symmetry Assignment | $\omega_{DFT}$ |
|------|------|------|------|------|------|------|------|
| | x=1 | x=0.7 | x=0.5 | x=0.3 | x=0 | | |
| J1 | | | | | 91.9±0.1 | | |
| S1 | | 108.0±0.0 | 102.3±1.1 | 103.0±0.0 | | | |
| P1 | 132.5±0.3 | 136.6±0.8 | 131.4±1.3 | 131.7±1.1 | | $B_g$ | 119 |
| S2 | | 158.0±0.0 | 153.9±1.2 | 153.3±1.1 | | | |
| P2 | 178.6±0.1 | 178.7±0.3 | 169.1±0.7 | 168.9±0.9 | 157.6±0.2 | $B_g$ | 157, 158 |
| S3 | | 210.9±1.5 | 203.6±0.9 | 206.0±1.0 | | | |
| S4 | | 222.2±0.2 | 217.9±0.1 | 219.7±0.1 | | | |
| P3 | 238.2±0.3 | 239.2±0.3 | 233.5±0.2 | 235.1±0.1 | 227.3±0.5 | $A_g$, $B_g$ | 215 |
| P4 | 255.1±0.1 | 256.9±0.1 | 251.9±0.2 | 253.2±0.1 | 248.7±0.5 | $A_g$ | 228, 229 |
| P5 | 281.7±0.3 | 284.3±0.9 | 280.6±0.8 | 283.2±0.8 | 280.2±0.1 | $A_g$, $B_g$ | 246, 247, 251 |
| P6 | 385.1±0.1 | 387.4±0.1 | 381.4±0.2 | 382.1±0.4 | 380.1±0.1 | $A_g$ | 350 |
| P7 | 441.8±0.6 | 448.1±0.4 | 438.9±0.2 | 442.5±0.3 | | $B_u$ | |
| J2 | | | | | 470.9±2.1 | | |
| P8 | 563.5±0.2 | 568.0±0.6 | 567.7±0.1 | 569.5±0.7 | 562.9±1.1 | $A_g$ | 504 |
| P9 | 589.6±0.1 | 591.7±0.1 | 586.0±0.1 | 587.5 ±0.1 | | $A_g$ | 526 |
| P10 | 599.0±0.5 | 601.5±0.1 | 597.0±0.0 | 598.6±0.1 | | $B_g$ | 527 |

**Table-S2:** BKT transition temperature (T$_{BKT}$) and antiferromagnetic transition temperature (T$_N$) for (Ni$_x$Fe$_{1-x}$)$_2$P$_2$S$_6$ series as extracted using our Raman data.

| X | $T_{BKT}$ | $T_N$ |
|---|------|------|
| 0 | 27 | 121 |
| 0.3 | 38 | 121 |
| 0.5 | 42 | 130 |
| 0.7 | 50 | 135 |
| 1 | 60 | 153 |

**Table-S3:** Atoms and the corresponding Wyckoff positions in the unit cell and irreducible representations of the phonon modes of monoclinic [point group - $C_{2h}$ and space group - $C2/m$ (#12)] (Ni/Fe)$_2$P$_2$S$_6$ at the $\Gamma$ point. Irreducible representation $\Gamma_{Total}$, $\Gamma_{Raman}$, $\Gamma_{IR}$, $\Gamma_{Acoustic}$ corresponds to total, Raman, infrared and acoustic phonon modes representation at the $\Gamma$ point, respectively. $R_{A_g}$ and $R_{B_g}$ are the Raman tensor for the Raman active $A_g$ and $B_g$ phonon mode, respectively.

| Atom | Type | Site | $\Gamma$ point phonon mode decomposition | Raman Tensors |
|------|------|------|------------------------------------------|---------------|
| Ni1/Fe1 | Ni/Fe | 4g | $A_g + A_u + 2B_g + 2B_u$ | $R_{A_g} = \begin{pmatrix} a & 0 & d \\ 0 & b & 0 \\ d & 0 & c \end{pmatrix}$ |
| P1 | P | 4i | $2A_g + A_u + B_g + 2B_u$ | |
| S1 | S | 4i | $2A_g + A_u + B_g + 2B_u$ | $R_{B_g} = \begin{pmatrix} 0 & e & 0 \\ e & 0 & f \\ 0 & f & 0 \end{pmatrix}$ |
| S2 | S | 8j | $3A_g + 3A_u + 3B_g + 3B_u$ | |
| $\Gamma_{Total} = 8A_g + 6A_u + 7B_g + 9B_u$, $\Gamma_{Raman} = 8A_g + 7B_g$, $\Gamma_{IR} = 5A_u + 7B_u$, $\Gamma_{Acoustic} = A_u + 2B_u$ | | | | |

**Table-S4:** List of the experimentally observed first ordered phonon modes at 4 K with 532 nm excitation laser for x = 1 & x = 0 and fitted parameters obtained by using anharmonic fitting model equation as described in the text. Units are in cm$^{-1}$.

|  | Mode | $\omega_0$ | C | D | $\Gamma_0$ | A | B |
|---|---|---|---|---|---|---|---|
| X=1 | P1 | 137.77 | -1.01 | 0.01 | 2.5+0.5 | 0.14+0.1 | 0.01+0.0 |
|  | P2 | 179.5±0.1 | -1.34±0.02 | -0.001±0.1 |  |  |  |
|  | P3 | 242.8+2.7 | -4.2+2.2 | 0.23+0.23 |  |  |  |
|  | P4 | 254.8+0.9 | 0.05+0.72 | -0.18+0.07 |  |  |  |
|  | P5 | 283.9±0.6 | -1.56±0.25 | -0.08±0.02 | 1.69±1.25 | 0.75±0.01 | 0.12±0.09 |
|  | P6 | 386.2+1.2 | -0.52+1.4 | -0.49+0.24 | 1.12 | 2.70 | 0.04+0.98 |
|  | P7 | 442+0.2 | -4.42+0.81 | -0.18+0.23 |  |  |  |
|  | P8 |  |  |  |  |  |  |
|  | P9 | 594.11+1.8 | -4.15+2.91 | -0.40+0.57 | 2.43+0.2 | 0.53+0.07 | 0.12+0.05 |
|  |  |  |  |  |  |  |  |
| X=0 | J1 |  |  |  |  |  |  |
|  | P2 | 157.4+0.4 | -0.60+0.29 | -0.03+0.02 | 5.49±2.0 | 0.50+0.05 | 0.001±0.07 |
|  | P3 | 227.9±0.12 | -1.0±0,02 | -0.12±0.01 |  |  |  |
|  | P4 | 247.6±1.3 | -0.09±0.01 | -0.14±0.10 | 2.19±1.46 | 0.51±0.29 | 0.49±0.10 |
|  | P5 | 280.7±0.9 | -1.06±0.81 | -0.07±0.09 | 4.90±0.17 | 0.17±0.06 | 0.05±0.01 |
|  | P6 | 380.7±0.9 | -0.89±1.16 | -0.19±0.17 | 3.69±1.05 | 0.64±0.27 | 0.06±0.30 |
|  | J2 |  |  |  |  |  |  |
|  | P8 |  |  |  |  |  |  |

**Table-S5:** List of the experimentally observed first ordered phonon modes at 4 K with 532 nm excitation laser for x = 0.7, 0.5 & 0.3, and fitted parameters obtained by using anharmonic fitting model equation as described in the text. Units are in cm$^{-1}$

| | Mode | $\omega_0$ | C | D | $\Gamma_0$ | A | B |
|---|---|---|---|---|---|---|---|
| | S1 | | | | | | |
| | P1 | 137.8±1.9 | -1.01±0.88 | 0.01±0.05 | | | |
| | S2 | | | | | | |
| | P2 | 178.5±1.3 | -0.77±0.77 | -0.02±0.06 | | | |
| | S3 | | | | | | |
| | S4 | 222.4±1.3 | -0.66±-96 | -0.08±0.09 | 8.48±0.21 | 0.73±0.08 | 0.01±0.01 |
| X=0.7 | P3 | 234.1±1.0 | 3.06±0.88 | -0.42±0.09 | | | |
| | P4 | 258.4±0.7 | -0.89±0.69 | -0.08±0.08 | | | |
| | P5 | 284.3±1.5 | 0.83±1.35 | -0.30±0.16 | 2.65±1.44 | 0.53±0.06 | 0.06±0.49 |
| | P6 | 389.1±0.9 | -2.01±1.21 | -0.12±0.19 | 5.58±5.43 | 0.63±0.05 | 0.04±0.02 |
| | P8 | 593.9±2.1 | -0.7±0.5 | -0.89±0.72 | 6.81±0.53 | 0.39±0.03 | 0.27±0.13 |
| | P9 | 593.9±2.1 | -0.73±0.12 | -0.88±0.72 | 7.0±3.71 | 0.40±0.05 | 0.09±0.72 |
| | P10 | 603.6±4.5 | -1.05±0.76 | -1.03±0.61 | | | |
| | | | | | | | |
| | S1 | | | | | | |
| | P1 | 134.6±1.9 | -1.20±0.82 | -0.003±0.04 | | | |
| | S2 | | | | | | |
| | P2 | 167.4±0.3 | -0.26±0.06 | -0.04±0.01 | | | |
| | S3 | | | | | | |
| | S4 | 218.8±0.2 | -1.45±0.14 | -0.03±0.01 | | | |
| X=0.5 | P3 | 230.7±1.7 | 1.46±0.64 | -0.29±0.06 | | | |
| | P4 | 246.9±1.1 | 4.65±1.11 | -0.77±0.14 | 1.02±0.22 | 0.26±0.07 | 0.16±0.05 |
| | P5 | 278.3±1.5 | 2.07±1.24 | -0.46±0.14 | 4.56±1.07 | 0.33±0.06 | 0.03±0.01 |
| | P6 | 382.1±0.1 | -0.64±0.01 | -0.28±0.02 | | | |
| | P8 | 571.0±9.8 | -1.13±0.08 | -1.55±0.19 | | | |
| | P9 | 589.5±0.3 | -1.12±0.05 | -1.09±0.06 | 5.45±0.72 | 2.12±0.11 | 0.14±0.04 |
| | P10 | 600.3±3.2 | -2.56±0.59 | -0.92±0.15 | 6.63±0.59 | 0.86±0.46 | 0.04±0.02 |
| | | | | | | | |
| | S1 | | | | | | |
| | P1 | 131.1±0.3 | -0.20±0.01 | -0.01±0.006 | | | |
| | S2 | | | | | | |
| | P2 | 166.7±2.2 | -1.98±1.26 | 0.12±0.09 | | | |
| | S3 | | | | | | |
| | S4 | 219.1±0.3 | -1.58±0.11 | -0.006±0.01 | 5.84±0.48 | 0.70±0.16 | 0.001±0.03 |
| X=0.3 | P3 | 233.9±0.31 | -0.97±0.10 | -0.04±0.02 | | | |
| | P4 | 252.6±0.3 | -1.02±0.14 | -0.001±0.01 | 0.96±0.12 | 0.76±0.30 | 0.07±0.05 |
| | P5 | 282.3±1.2 | -1.31±1.11 | -0.01±0.13 | | | |
| | P6 | 383.2±0.7 | -2.51±0.87 | 0.14±0.13 | | | |
| | P8 | 579.6±16.9 | -19.02±2.3 | 2.70±5.73 | | | |
| | P9 | 588.8±3.0 | -2.18±0.15 | -0.39±0.17 | 6.21±0.21 | 1.16±0.12 | 0.34±0.03 |
| | P10 | 600.0±3.4 | -2.85±0.91 | -0.53±0.23 | 5.43±4.27 | 2.52±0.22 | -0.42±1.49 |